\documentclass[]{aastex631}

\usepackage{epsfig} 
\usepackage{natbib}
\usepackage{amsmath}
\usepackage{bm} 

\renewcommand{\vec}[1]{\bm{#1}} 

\newcommand{\pder}[2]{\frac{\partial #1}{\partial #2} }
\newcommand{\grad}{ {\bf \nabla } }
\newcommand{\divv}{ {\bf \nabla} \cdot }
\newcommand{\curl}{ {\bf \nabla} \times}

\def\edit#1{{\color{red}{#1}}}

\shorttitle{Observations and Modeling of QFP Waves}
\shortauthors{Wang et al.}

\graphicspath{{./}{figures/}}

\begin{document}


\title{Quasi-Periodic Fast-Mode Wave Trains Associated with the 2015-Jun-22 M6.5 Flare in AR~12371: Observations and 3D MHD Modeling}

 \correspondingauthor{Tongjiang Wang}
 \email{tongjiang.wang@nasa.gov}

\author[0000-0003-0053-1146]{Tongjiang Wang}
\affiliation{Institute for Astrophysics and Computational Sciences, Catholic University of America, 620 Michigan Avenue NE,
 Washington, DC 20064, USA}
\affiliation{NASA Goddard Space Flight Center, Code 671, Greenbelt, MD 20770, USA}

\author[0000-0001-8794-3420]{Wei Liu}
\affiliation{Bay Area Environmental Research Institute, NASA Research Park, Building 18, Mailstop 18-4, Moffett Field, CA 94035-0001, USA}
\affiliation{Lockheed Martin Solar and Astrophysics Laboratory, 3251 Hanover Street, Palo Alto, CA 94306, USA}

\author[0000-0003-0602-6693]{Leon Ofman}
\affiliation{Institute for Astrophysics and Computational Sciences, Catholic University of America, 620 Michigan Avenue NE,
 Washington, DC 20064, USA}
\affiliation{NASA Goddard Space Flight Center, Code 671, Greenbelt, MD 20770, USA}

\author[0000-0003-4043-616X]{Xudong Sun}
\affiliation{Institute for Astronomy, University of Hawai‘i at M\={a}noa,  Pukalani, HI 96768, USA }

\author[0000-0002-9672-3873]{Meng Jin}
\affiliation{Lockheed Martin Solar and Astrophysics Laboratory, 3251 Hanover Street, Palo Alto, CA 94306, USA}

\begin{abstract}
Quasi-periodic fast-propagating (QFP) wave trains are a distinctive form of magnetohydrodynamic disturbance frequently observed in the solar corona. Yet their excitation mechanism and propagation characteristics are not well understood. In this study, we investigate a well-observed QFP wave event associated with an M6.5-class flare and coronal mass ejection that occurred in active region (AR) 12371 on 2015 June 22 by combining multi-wavelength observations from SDO/AIA and HMI with data-inspired 3D MHD simulations. The QFP wave trains propagating at high speeds of 1140$-$1760 km~s$^{-1}$ are detected in the AIA 171 \AA\ channel, following  global EUV wave fronts visible at 171 and 193~\AA\ traveling at considerably lower speeds of 300$-$510 km~s$^{-1}$.
 Wavelet analysis reveals consistent 2--4 minutes  periodicities in both the QFPs and flare quasi-periodic pulsations (QPPs) observed in UV/EUV and hard X-ray emissions, suggesting a common origin likely linked to intermittent magnetic reconnection. Guided by these observations, we construct realistic 3D MHD models incorporating dense fan-loop structures and periodic drivers applied at different locations. The simulations reproduce the key characteristics of the observed wave trains. Comparison between cases with and without a coronal background (non-loop plasma emission) indicates that coronal density structuring significantly modifies the detected wave amplitudes and propagation patterns. Our results highlight the importance of realistic coronal magnetic configurations in modeling QFP dynamics and suggest that their observed association with fan loops in AIA 171 \AA\ may represent a temperature-dependent visibility effect rather than a genuine confinement of the waves.

\end{abstract}


\section{Introduction} \label{sec:intro}


Extreme-ultraviolet (EUV) waves in various forms are manifestations of magnetohydrodynamic (MHD) waves in the magnetized solar coronal plasma and can serve as powerful tools for coronal seismology and flare diagnostics \citep[e.g.][]{wan16,LO14,nak24}. Among these phenomena, quasi-periodic fast-propagating (QFP) wave trains, discovered with SDO/AIA \citep{Liu10,liu11,liu12}, stand out as high-speed, arc-shaped fronts that emanate from flare sites and travel 
in the structured coronal media, often along a funnel-shaped path.  
Understood as fast-mode magnetosonic waves, QFPs exhibit a broad range of phase speeds (from a few hundred to a few thousand km~s$^{-1}$), periods from sub-minute to several minutes, small amplitudes typically on the 1\%$-$5\% level in EUV intensity, and a strong tendency to appear in EUV channels sensitive to relatively cool coronal plasma (notably AIA 171 \AA), although their detectability depends sensitively on instrument response and background emission due to the small signal-to-noise of these events. QFP waves are best seen in difference images or running difference animations. Reviews and surveys summarize both the rapidly growing observational phenomenology and the diagnostic potential of QFPs for probing coronal magnetic field and plasma properties \citep{LO14,liuw16,shen22}.

A commonly observed feature of QFP events is their temporal correlation with flare quasi-periodic pulsations (QPPs).  Although it remains challenging to directly distinguish among the various proposed QPP mechanisms -- primarily including MHD waves, wave-modulated energy release, and busty quasi-periodic magnetic reconnection \citep[e.g.,][]{zim21,Ashf26,real26}, many observations show that the onset of QFP wave trains and their dominant periodicities closely track oscillatory characteristics in flare light curves across multiple wavelengths. This correspondence suggests a common driver or tight coupling between the flare energy-release process and the propagating fast waves \citep[e.g.,][] {liu11,liu12,shen12,shen13,shen18a,miao21,zhou21}. 
At the same time, such correlations are not always present: 
several studies report QFPs without clear co-temporal QPPs in the flare light curves \citep[e.g.,][]{shen18b,miao19}, and vice versa, indicating that multiple excitation pathways may operate under different coronal conditions. Consequently, disentangling driver mechanisms requires joint space-time analysis 
and knowledge of the 3D magnetic geometry in each event.

The generation mechanism of QFP wave trains remains under active debate. Leading scenarios include (1) periodic reconnection or plasmoid formation in flare current sheets that periodically produce compressional disturbances \citep{yang15,Downs15,tak16,mond24,liak25}, which explains the above noted QFP-QPP correlation, (2) impulsive footpoint drivers launching broadband pulses that evolve via waveguide dispersion into multiple fronts \citep{pas13,pas17,nist14,kolot21,shi25,shi26}, and (3) driver effects such as CME flank expansions, breakout eruptions, or jet-loop interaction \citep{zhang15,shen18c,mei20,wang21,hu24}. Each mechanism offers specific predictions for periodicity, propagation direction, wave amplitude, and visibility in EUV channels.  For example, a recent study by \citet{mond25}, based on 2D MHD simulations, demonstrated that impulsive magnetic reconnection in a flare current sheet can excite surface sausage-mode oscillations and their subsequent propagation and leakage, suggesting that the periodicity of QFP waves may be closely linked to oscillatory dynamics of the current sheet. However, no single model yet reproduces the full range of observed behaviors, warranting the need for integrated observational and modeling efforts, as we aim to do in this paper.

 Most QFP events are associated with global EUV waves and CMEs \citep[e.g.,][]{liuw16}. Global EUV waves (also known as ``EIT waves" or large-scale EUV waves) represent a distinct manifestation of fast magnetoacoustic disturbances in the corona, differing from the localized QFP wave trains primarily in their excitation mechanisms, spatial scales, and observational characteristics. Global EUV waves are typically large-scale, large-amplitude (5\%--35\% intensity increase)  single or broadband pulses driven by impulsive eruptions such as CMEs, often exhibiting nonlinear behavior (e.g., wave steepening) and complex multi-component fronts that may include both fast and slower signatures \citep[e.g. reviews by][]{LO14, warm15, chen16, shen22,zhen24}. In contrast, QFP wave trains consist of narrow-band, quasi-periodic wave packets that are commonly observed along coronal funnel-like structures, with substantially smaller amplitudes. They are most frequently observed during the impulsive phase of flares, but can also appear after the flare peak. Although both phenomena may occur during the same eruptive event and involve fast-mode wave physics, their differing spatial, morphological, and temporal characteristics suggest that they arise from different physical processes.


Numerical MHD modeling of QFP wave trains has achieved significant progress in recent years. Numerous 2D and 2.5D simulations have successfully reproduced high-speed fast-mode waves or pulses, either excited by a spatially localized impulsive driver within a coronal waveguide \citep[e.g.,][]{pas13,pas17}, or generated spontaneously by bursty magnetic reconnection in flare current sheets \citep[e.g.,][]{yang15,liak25}. However, the idealized magnetic geometries adopted in these models and the omission of fully 3D AR magnetic complexity and coronal inhomogeneity prevent direct quantitative comparison of the simulated results with observations. Based on a 3D dipolar AR magnetic field model that includes gravitational stratification, \citet{ofm11} excited fast-mode waves using periodic velocity perturbations and found that the modeled wave characteristics were roughly consistent with observations, providing the first modeling-supported interpretation in terms of fast mode waves (see, also \citealt{ofm18}, which models counter-propagating QFPs observed in dual flare event). Recently, \citet{ofm25} extended this work by incorporating a more realistic magnetic configuration derived from potential-field extrapolations of observed photospheric magnetograms, demonstrating that the resulting wave properties show better agreement with observations than those from earlier idealized models, such as  structure (confinement) and damping properties of the modeled QFP wave train. Modeling studies show that QFP waves can exhibit nonlinear effects such as wave front steepening and nonlinearity-generated slow mode waves \citep[e.g][]{ofm18}. However, simulated QFPs do not form shocks, consistent with observations.

Despite these advances, realistic three-dimensional MHD simulations directly constrained by specific observations remain scarce. Most existing models focus on simplified magnetic configurations or assume idealized driver geometries, limiting their ability to capture the true magnetic complexity, thermodynamic stratification, and background emission conditions of active regions. Consequently, important observational properties, such as the apparent association of QFP wave trains with fan loops, the variation of propagation speeds with distance, and the dependence of observed amplitudes on coronal temperature response, are yet to be fully understood in a unified framework. Motivated by these challenges and building upon the previous model of \citet{ofm25}, we perform realistic 3D MHD simulations of QFP wave trains in NOAA AR 12371 on 2025 June 22, using an initial magnetic configuration that incorporates dense coronal loop structures and is guided by detailed observational constraints from SDO/AIA. By comparing the simulated and observed wave characteristics, we aim to identify the physical origin and excitation mechanisms of the QFPs, assess the influence of coronal background emission (from non-loop plasma) on the observed wave patterns and their visibility in AIA channels, and provide improved diagnostics of coronal wave energetics and propagation based on self-consistent 3D numerical modeling. 

The paper is organized as follows. Section~\ref{sec:obs} presents the analysis of the observed QFP event and summarizes the key results that motivate our modeling study. Section~\ref{sec:model} describes the 3D MHD numerical model, including the initial conditions and the setup of the periodic drivers. Section~\ref{sec:result} presents the simulation results for two models with different driver configurations. Finally, Section~\ref{sec:dac} summarizes the main findings and provides a discussion of their physical implications.

\section{Observations} 
\label{sec:obs}

In this study, we analyze a QFP wave train event on 2015 June 22, associated with a CME and a GOES M6.5-class flare in AR 12371. This flare was observed in exquisite detail by the 1.6~m New Solar Telescope at Big Bear Solar Observatory \citep{Jing16} and found to exhibit slipping magnetic reconnection along a dimming channel \citep{Jing17}.

\begin{figure}
\centering
\includegraphics[width=0.7\linewidth]{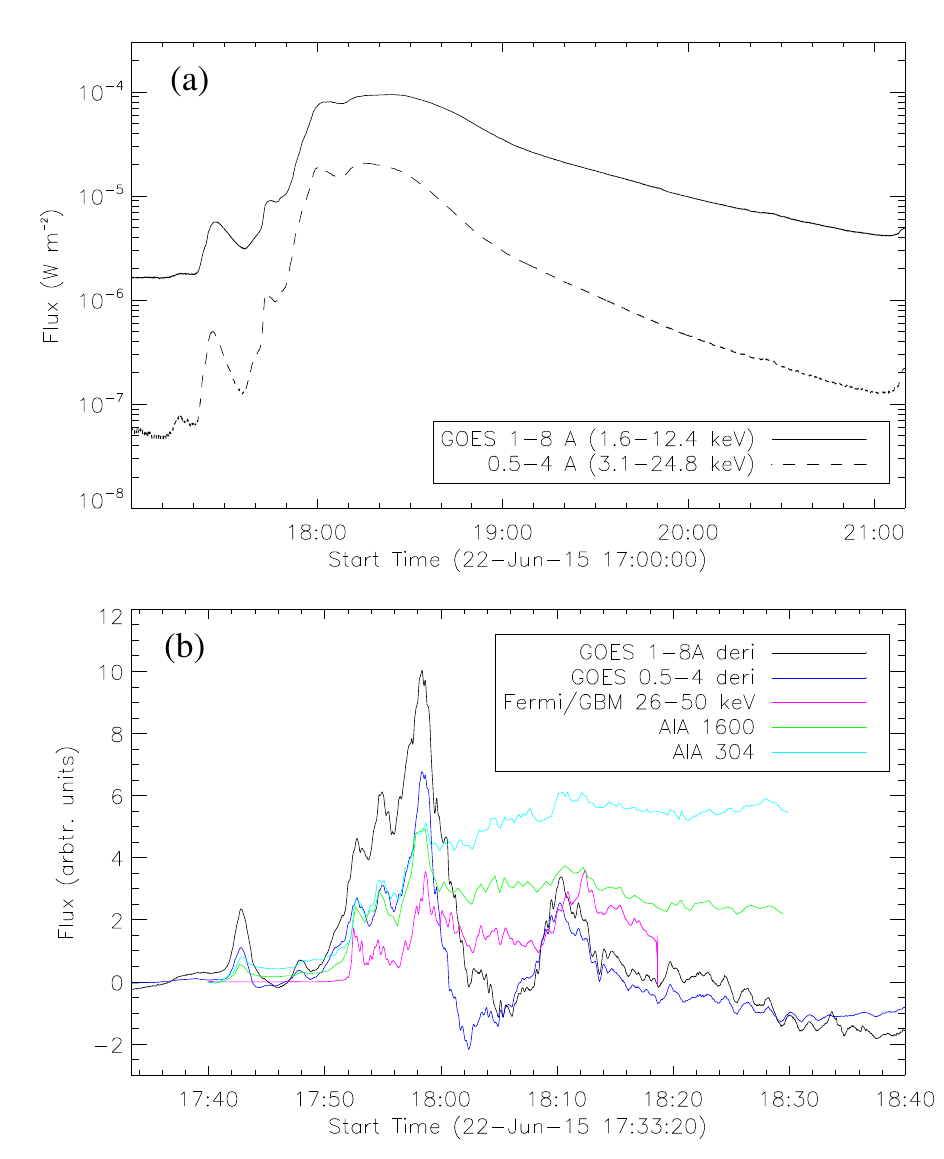}
   \caption{(a) GOES soft X-ray fluxes and (b) their time derivatives (smoothed with a 1-minute boxcar) in comparison with the Fermi/GBM 26–50 keV X-ray flux (smoothed with a 12-second boxcar), together with light curves in SDO/AIA 1600 and 304 \AA. 
\label{fig:lc}}
\end{figure}

\begin{figure*}
\centering
\includegraphics[width=0.9\linewidth]{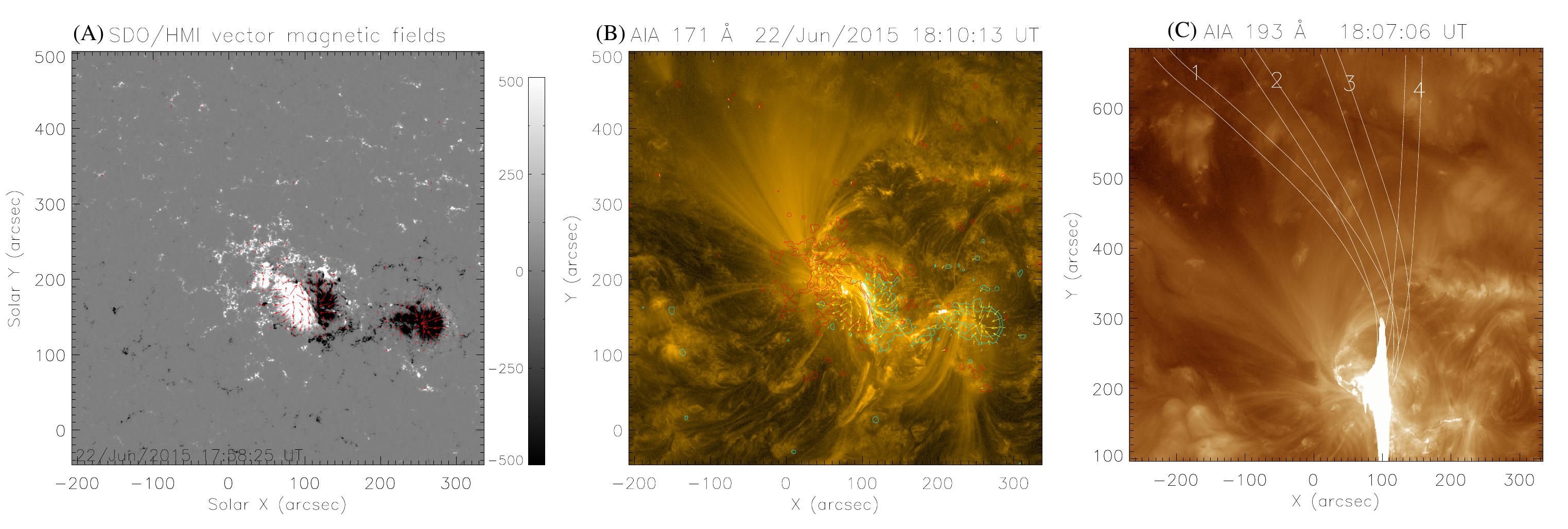}\\
\includegraphics[width=0.9\linewidth]{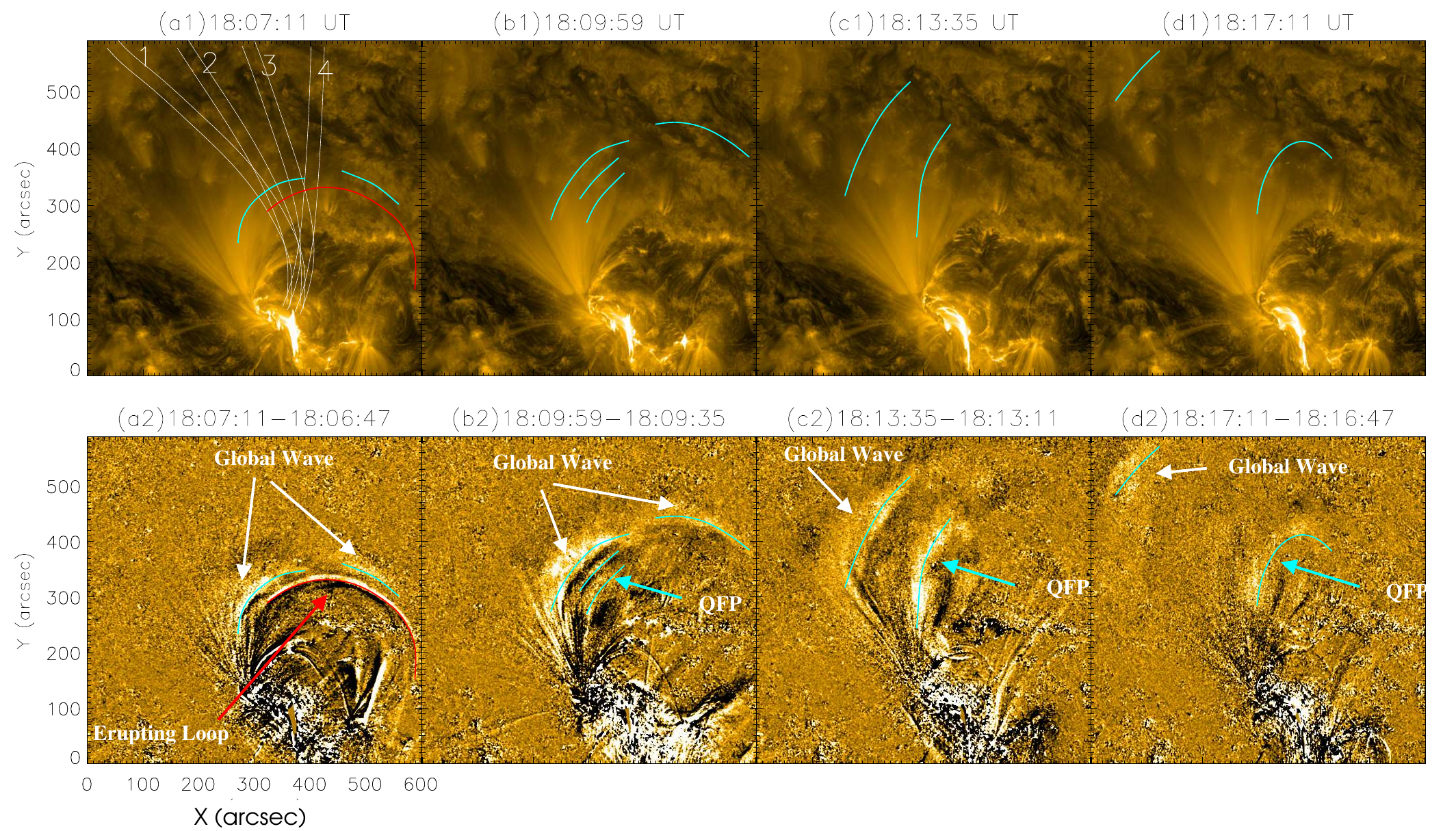}
        \caption{\textit{Upper panels:} (A) SDO/HMI vector magnetogram of AR 12371 observed at 17:58:25~UT on 2015 June 22. The background shows the longitudinal magnetic field component scaled between $\pm$500~G, with positive polarity shown in white and negative in black.
        (B) SDO/AIA 171~\AA\ image showing flaring loops 
        near the magnetic neutral line. The overlaid contours represent the vertical ($B_z$) component of the magnetic field at levels of $\pm$500 and $\pm$100~G, with red indicating positive and cyan negative polarities. The same arrows in (A) and (B) denote the transverse magnetic field vectors, with strengths ranging from 100 to 2100~G.
        (C) AIA 193~\AA\ image taken at 18:07:06~UT.
        \textit{Middle and bottom panels:} AIA 171~\AA\ direct (\textit{middle}) and running difference (\textit{bottom}) images showing the flare-generated QFP wave trains. The red curve outlines the expanding flux rope, while the cyan curves trace the propagating wave fronts. The four numbered slices, shown in white in (C) and (a1), are used to produce the time--distance plots shown in Figure~\ref{fig:tdismap}. 
          An animation corresponding to the middle and bottom panels of this figure is available. The accelerated animation (12 s duration) presents the time sequence of AIA 171 \AA\ images (top panel) and their 24 s running-difference (bottom panel) covering the interval from 18:00:47 to 18:29:59 UT on 2015 June 22. (An animation of this figure is available in the online article.)
\label{fig:bimg}}
\end{figure*}

The flare started at 17:39 UT and peaked at 18:23 UT. Figure~\ref{fig:lc} presents the temporal variations of the GOES soft X-ray flux in the 1$-$8 \AA\ and 0.5$-$4 \AA\ bands (Panel a), together with comparisons of their time derivatives, the Fermi/GBM hard X-ray flux in 26$-$50 keV, and the flare kernel light curves observed in the AIA 1600 and 304 \AA\ channels (Panel b). To reduce noise, the GOES derivatives and the Fermi/GBM curves were smoothed with a 1-min and 12-s boxcar, respectively. The AIA light curves were obtained by averaging the intensity over the region X=[0, 300] arcsec and Y=[0, 400] arcsec (see Figure~\ref{fig:bimg}B). A clear correspondence of peaks is seen between 17:50 and 18:00 (Figure~\ref{fig:lc}b). This behavior is expected, since the GOES flux derivative serves as a proxy for the HXR light curve during the impulsive phase \citep{den93}, while the AIA 1600 and 304 \AA\ emissions represent the prompt response of the lower atmosphere to impulsive energy deposition \citep[e.g.,][]{qiu12}.

Figure~\ref{fig:bimg}(A) shows the vector magnetogram of AR 12371 observed by SDO/HMI at 17:58 UT, and Figure~\ref{fig:bimg}(B) presents the AIA 171 \AA\ image at 18:10 UT, overlaid with the vector magnetic fields from panel (A). The main flare ribbons are located approximately along the magnetic neutral line. Figure~\ref{fig:bimg}(C) displays the AIA 193 \AA\ image at 18:07 UT, where four narrow slices (about 25$^{''}$ wide), originating from the flare source, are marked approximately along the wave propagation directions. These slices are used to construct time-distance maps for measuring wave propagation in different directions.

The bottom two rows of Figure~\ref{fig:bimg} illustrate the evolution of the QFP wave trains in the AIA 171 \AA\ channel (panels a1--d1) and their running-difference counterparts (panels a2--d2). Following the flare ribbon brightening, a slowly expanding loop structure (likely an erupting flux rope) was observed moving northward,  generating initial global (or large-scale) EUV wave fronts at 18:07 UT (marked ``Global Wave'' in panel a2). The EUV wave fronts propagated toward the northeast and north (panels a2--d2). Between 18:10 and 18:20 UT, several QFP wave trains appeared following the EUV fronts, most clearly seen against the fan loops (see panels b1--d1 and b2--d2). These fronts and wave trains are more distinct in the running-difference images. Animations associated with the bottom two rows of Figure~\ref{fig:bimg} are provided in the online version. Notably, the initial EUV wave fronts are visible in both the AIA 171 and 193 \AA\ channels, whereas the QFP wave trains are detected only in 171 \AA.

\begin{figure*}
\gridline{\fig{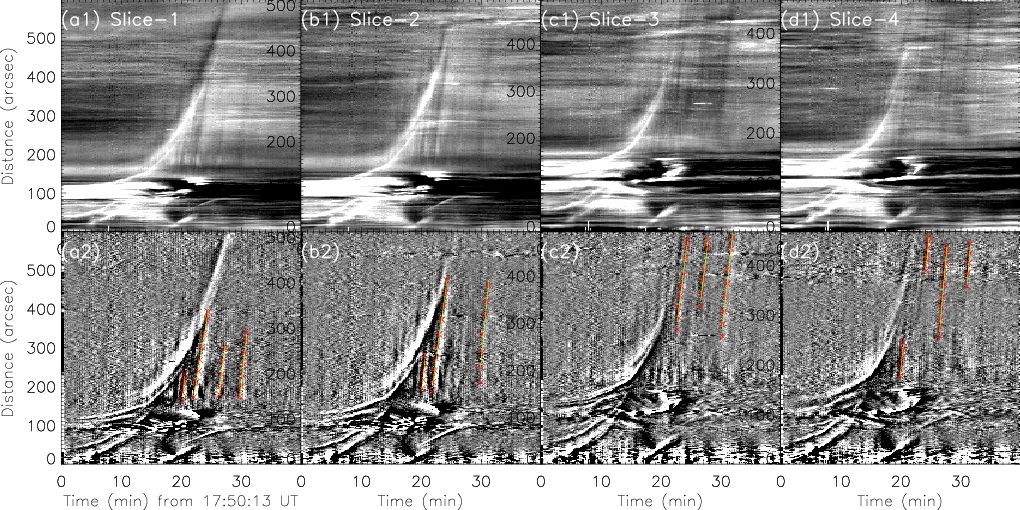}{1.0\textwidth}{}}
\gridline{\fig{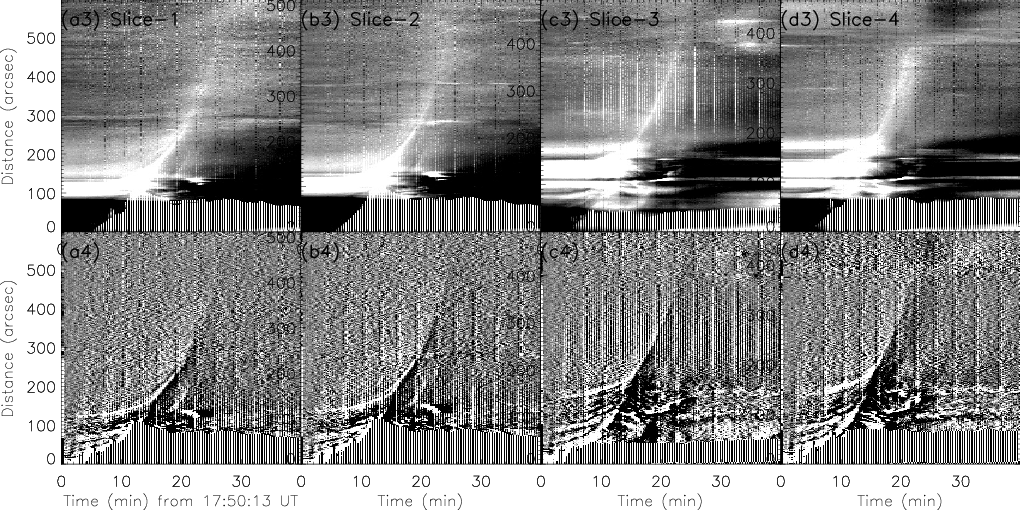}{1.0\textwidth}{}}
\caption{\textit{Top two rows:} (a1)-(d1) Time--distance maps of the AIA 171~\AA\ relative intensity, defined as $(I(t)-I_0)/I_0$, where $I_0$ denotes the average intensity over time at each position along the slice. (a2)-(d2) Same as the upper panels, but for the running-difference images. The cross symbols and slanted lines mark the identified wave features and their linear fits used to measure the propagation speeds. \textit{Bottom two rows:} (a3)-(d3) Time--distance maps of the AIA 193~\AA\ emission in base-difference form. (a4)-(d4) Same as (a3)-(d3), but for the running-difference images. The alternating vertical dark and bright strips are artifacts resulting from variations in the image exposure times due to AIA's automatic exposure control (AEC) during flares. 
\label{fig:tdismap}}
\end{figure*}

The top two rows of Figure~\ref{fig:tdismap} present the time-distance maps along the four slices for the background-subtracted 171 \AA\ intensity (panels a1--d1) and their running differences (panels a2--d2). The initial, slowly accelerating flux rope, originating at a distance of about 120$^{''}$, is clearly visible. The global EUV wave fronts, generated around 180$-$200$^{''}$, are evident in panels (a1)--(c1) and propagate with speeds of 320$-$510 km~s$^{-1}$. Notably, the bright wave front along Slice 1 evolves into a dimmed feature at about 400$^{''}$, likely due to heating by adiabatic compression \citep[e.g.,][]{liu12,van15,jin22}. The subsequent QFP wave trains appear mainly as dark, slanted streaks, propagating at much higher speeds than the global  EUV wave fronts. The wave trains along Slices 1 and 2 originate near 180$^{''}$, whereas those along Slices 3 and 4 begin farther out, at about 280$^{''}$. 

By applying linear fits to 14 QFP wave features identified in Figure~\ref{fig:tdismap}(a2)--(d2), we obtained propagation speeds ranging from 1142 to 1755 km~s$^{-1}$, with a mean value of $1512\pm226$ km~s$^{-1}$. Thus, the QFP wave trains propagate at more than three times the speed of the initial EUV wave fronts.

The bottom two rows of Figure~\ref{fig:tdismap} display the time-distance maps from AIA 193 \AA\ images and their running differences along the same four slices as in 171 \AA. In this channel, only the slowly accelerating flux rope and the global EUV wave fronts are detected, while the QFP wave trains are absent.

\begin{figure*}
\plotone{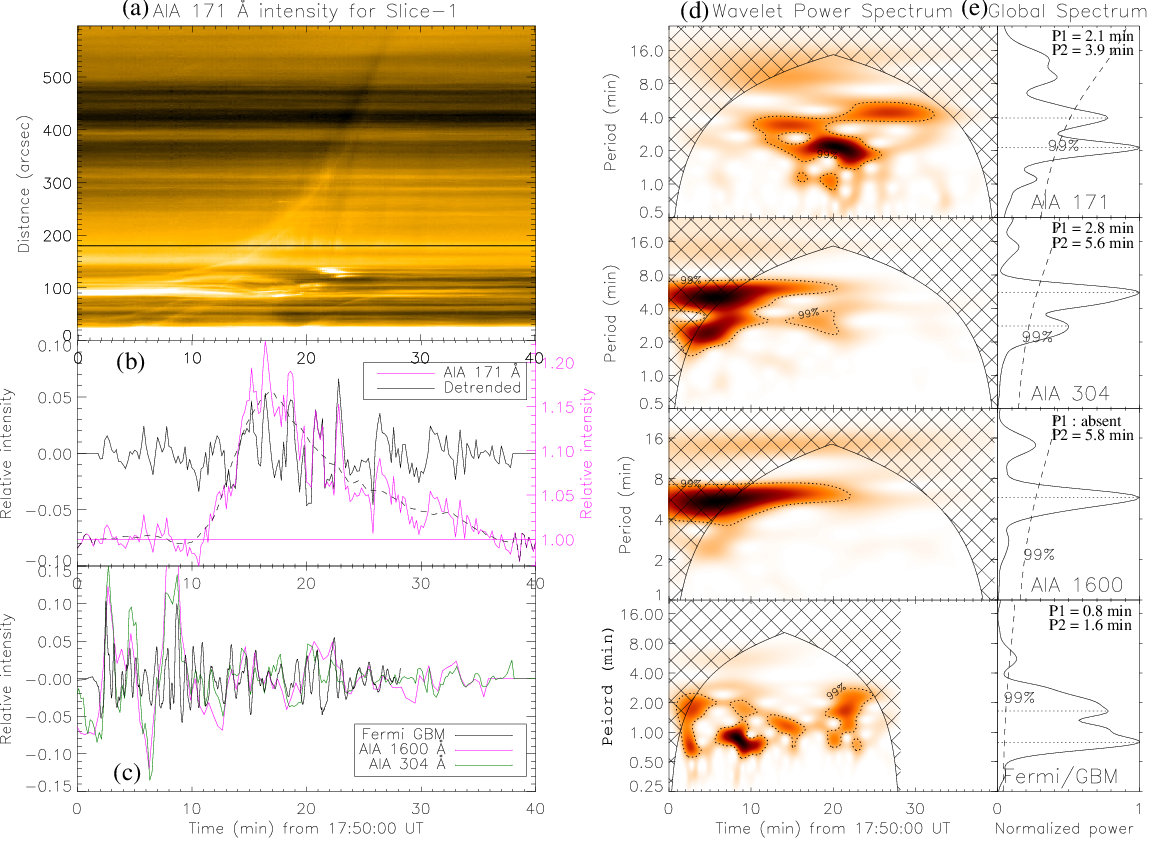}	 
\caption{(a) Time--distance map of AIA 171~\AA\ intensity along Slice~1 (in the logarithmic scale).  (b) Time profiles of AIA 171~\AA\ intensity (purple solid line), shown relative to the averaged pre-event background (purple horizontal line) and its detrended signal (black solid line) along a spatial cut indicated by the black line in panel~(a). The black dashed line shows the background trend. (c) Comparison of background-removed Fermi/GBM 26--50~keV X-ray flux (amplitudes reduced by a factor of~7) with background-removed relative intensities of AIA 1600 and 304~\AA. Those detrended light curves shown in (b) and (c) are calculated as $(I(t)-I_{\rm bg}(t))/I_{\rm bg}(t)$, where, $I_{\rm bg}(t)$ represents a slowly varying background. (d) Wavelet power spectra of the intensity fluctuations in AIA 171~\AA, 304~\AA, 1600~\AA, and Fermi/GBM X-rays (from top to bottom). Dark colors indicate regions of high power, and the dotted contours enclose areas exceeding the 99\% confidence level. The black cross-hatched region marks the cone of influence where period estimates become unreliable. (e) Corresponding global wavelet power spectra (solid lines). Peaks above the 99\% confidence level (dashed line) are statistically significant.  
\label{fig:wlet}}
\end{figure*}

To investigate the origin of the QFP wave trains observed in AIA 171 \AA, we analyzed their periodicities and compared them with those of the flare light curves in AIA 1600 \AA, AIA 304 \AA, and Fermi/GBM hard X-rays using the wavelet method \citep{torr98}. The 171 \AA\ light curve was extracted at a distance of 180$^{''}$ along Slice 1, where the QFP trains originated (Figure~\ref{fig:wlet}a). The 1600 and 304 \AA\ light curves were obtained as the mean intensity over the flare kernels. The 171 and 304 \AA\ images have a cadence of 12 s, while the 1600 \AA\ images have 24 s. The GBM hard X-ray flux was recorded at 0.064 s cadence during the flare. To match the AIA cadences, we smoothed the GBM flux with a 12 s boxcar and interpolated it to a uniform 5 s interval. For the wavelet analysis, the AIA light curves were detrended by subtracting a 20-frame boxcar, and the GBM flux was detrended by subtracting its 64 s smoothed curve.

 Figure~\ref{fig:wlet}b shows the light curve of the QFP wave trains in AIA 171 \AA\ and its detrended signal (with amplitudes of $\sim2-4\%$). Note that the time profile of the background trend (black dashed line) exhibits behavior consistent with the typical global EUV wave front \citep[e.g.,][]{Liu10,mann23}. Its amplitude is about $17\%$ of the pre-event background, which is substantially larger than that of the QFP waves, indicating that the large-scale background variation is dominated by the global EUV disturbance. Figure~\ref{fig:wlet}c presents those of the flare fluxes from Fermi/GBM and AIA 1600 and 304 \AA. The flare light curves exhibit correlated fluctuations (i.e., QPPs) between the hard X-rays and AIA UV/EUV bands, except for the shorter-timescale peaks that appear only in the hard X-rays. The corresponding wavelet power spectra are shown in Figure~\ref{fig:wlet}d. Strong fluctuations in the flare light curves mainly occur during the rising phase (before 18:00 UT), whereas the strongest QFP wave trains appear near the GOES soft X-ray peak (18:00–18:20 UT).  The global spectral analysis reveals two dominant periodic components (labeled P1 and P2 in Figure~\ref{fig:wlet}e): 2.1 and 3.9 minutes in AIA 171 \AA, 2.8 and 5.6 minutes in AIA 304 \AA, N/A and 5.8 minutes in AIA 1600 \AA, and 0.8 and 1.6 minutes in Fermi/GBM 26$-$50 keV. The absence of the shorter-period component (P1) in the AIA 1600 \AA\ data is likely due to its lower cadence (24~s) relative to the other channels (12~s). These results suggest that the periodicities of the QFP wave trains are consistent with those of the flare light curves, supporting their origin in the flare region.

We summarize the observed characteristics of the QFP wave trains in AR 12371, which motivate our MHD simulations, as follows:
\begin{enumerate}
\item The QFP wave trains appear after the initial global EUV wave fronts. The QFPs are detected only in AIA 171 \AA, whereas the global EUV wave fronts are visible in both AIA 171 and 193 \AA.

\item The global EUV wave fronts emerge ahead of an accelerating flux rope, appearing first as bright disturbances in 171 \AA\ and later evolving into dark features. The QFP wave trains originate at a distance of about 133 Mm from the flare source, mainly as dark features associated with fan loops in 171 \AA.

\item The global EUV wave fronts show gradual acceleration, propagating at 300--510 km~s$^{-1}$, while the QFP wave trains travel at nearly constant speeds of $1512\pm226$ km~s$^{-1}$, with relative amplitudes of 2--4\%.

\item Wavelet analysis reveals periodicities of 2.1 and 3.9 minutes for the QFP wave trains in AIA 171 \AA. Their approximate match to the flare flux fluctuations (about 1--6 minutes) suggests that the QFP wave trains originate from the flare source.
\end{enumerate}

 \begin{figure*}
\plottwo{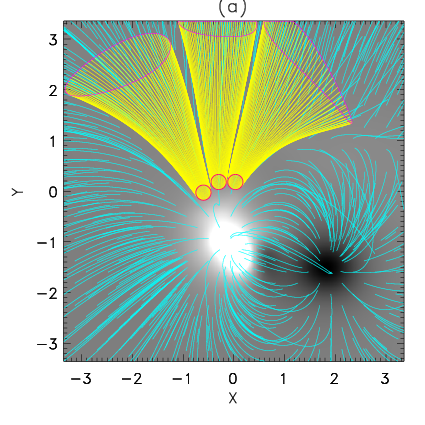}{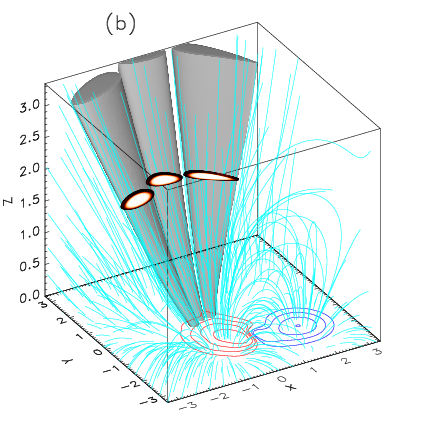}
\gridline{\fig{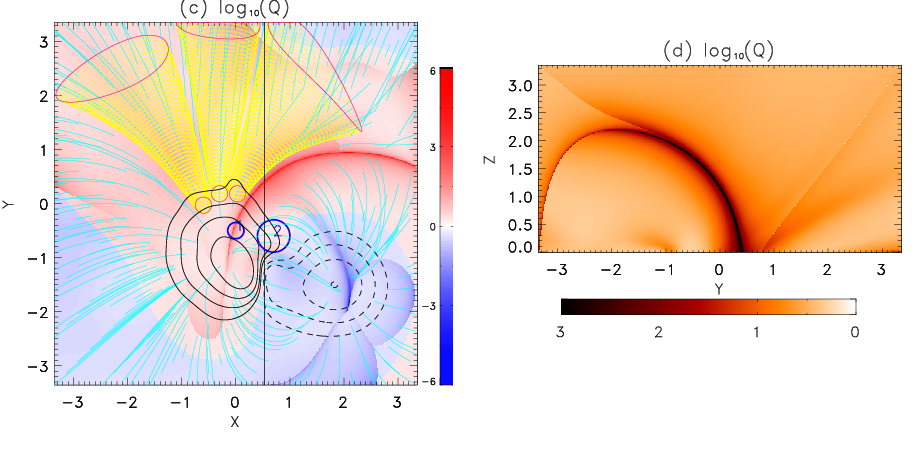}{0.9\textwidth}{}}
\caption{\textit{Top panels:} (a) Initial 3D potential field model of AR~12371 including three dense loops constructed by tracing magnetic field lines (yellow) to simulate the AIA fan loops. Note that the $xyz$ coordinates are in dimensionless unit. The background plane at $z=z_{\rm min}$ shows the radial magnetic field component at a height of 17~Mm above the photosphere, scaled between $\pm100$~G, with positive and negative polarities represented in white and black, respectively. Cyan lines denote additional field lines extrapolated from the bottom boundary. The yellow lines correspond to field lines defining the three loop models, whose footpoints at $z=z_{\rm min}$ are marked by small red circles and endpoints at $z=z_{\rm max}$ by pink circles. (b) 3D view of the initial magnetic configuration showing the three loops with enhanced density having a sharp Gaussian cross-sectional profile. The isosurface represents a density contrast of 1.01. The cross-section at $z=1.7$ illustrates the density contrast distribution $\chi_\rho$ ranging from~1 to~2. The contours show the $z$-component of the magnetic field at $z=z_{\rm min}$ with levels of $\pm12$, $\pm25$, $\pm50$, and $\pm100$~G, where red and blue denote the positive and negative polarities, respectively. The cyan lines have the same meaning as in (a).  \textit{Bottom panels:} Squashing factor ($Q$) maps calculated from the potential field model. (c) Logarithmic distribution of the squashing factor ($\mathrm{sign}(B_z)\log_{10}(Q\ge2)$) in the $xy$-plane at $z=z_{\rm min}$. Positive and negative magnetic polarities are shown in red and blue, respectively, with bright regions indicating high $Q$ values. Solid and dashed contours denote the $z$-component of the magnetic field with levels of $\pm12$, $\pm25$, $\pm50$, and $\pm100$~G. The two thick blue circles mark the locations of the wave drivers in Model~1 (smaller circle) and Model~2 (larger circle). Other lines are as described in (a). (d) Logarithmic squashing factor ($\log_{10}(Q\ge2)$) in the $yz$-plane at $x=0.53$ (indicated by the vertical line in (c)). High-$Q$ regions are shown in dark shading. 
 \label{fig:lpmd}}
 \end{figure*}

\section{MHD Model}
\label{sec:model}
While AIA EUV images with high cadence and resolution are invaluable for identifying QFP wave trains, they are inherently limited by observational challenges such as line-of-sight (LOS) projection effects, instrumental effects in temperature response, and unknown density and magnetic field distributions. These constraints often prevent a definitive identification of MHD wave modes and their interactions. To overcome these limitations, 3D MHD modeling in a realistic magnetic configuration becomes crucial. It allows us to disentangle the physical factors influencing observed features, clarify the nature of MHD wave modes, and investigate the effects of varying wave sources by comparing synthetic observations with real data.

For this purpose, we solve the nonlinear resistive 3D MHD equations using the NLRAT code, which was specifically developed to study wave dynamics in coronal ARs and loops \citep[e.g.,][]{OT02,mcl08,ofm11,prov18,ofm18} and recently extended to study QFP waves in a realistic coronal magnetic field by \citet{ofm25}. The MHD equations account for the effects of gravity, compressive viscosity, heat conduction, and optically thin radiation, expressed in a flux-conservative and dimensionless form \citep[see details in][]{ofm22,wan24}:

\begin{equation}
 \pder{\rho}{t} + \divv ( \rho \vec{V} ) =0,  \label{eq:cn}
\end{equation}
\begin{equation}
 \pder{(\rho \vec{V})}{t} + \divv \left[ \rho \vec{V}\vec{V} + \left( E_u \, p + \frac{\vec{B} \cdot \vec{B}}{2} \right) \vec{I} - \vec{B} \vec{B} \right] = \frac{1}{F_r} \rho \vec{F_g} +\vec{F_v}, \label{eq:mm}
\end{equation}
\begin{multline}
  \pder{(\rho E)}{t} + \divv \left[\vec{V} \left(\rho E + E_u \, p + \frac{\vec{B} \cdot \vec{B}}{2} \right) - \vec{B}(\vec{B} \cdot \vec{V}) + \frac{1}{S} \curl \vec{B}\times\vec{B} + \vec{V}\cdot\mathbf{\Pi} \right] \\
 =\frac{1}{F_r} \rho \vec{F_g} \cdot \vec{V} + \grad_\| (\kappa_\|\grad_{\|}T) - Q_{\rm rad} + H_{\rm in},
         \label{eq:en}
\end{multline}
\begin{equation}
  \pder{\vec{B}}{t} = \curl (\vec{V}\times\vec{B}) + \frac{1}{S} \nabla^2 \vec{B}. \label{eq:ind}
\end{equation}
In the above equations, the total energy density is represented as $\rho{E}=\frac{E_u p}{(\gamma-1)}+\frac{\rho{V^2}}{2}+\frac{B^2}{2}$, where $p$, $\rho$, $V$, and $B$ correspond to the dimensionless pressure, density, velocity, and magnetic field, respectively. The detailed normalization parameters are listed in Table~\ref{tab:par}. The operator $\grad_\|$ represents the $B$-parallel gradient, defined as $\grad_\|=\frac{1}{|B|}\vec{B}\cdot\grad$. $\vec{F_g}=-\frac{1}{(10+z-z_{\rm min})^2}\vec{e_z}$ is the gravity term. $\kappa_\|=7.8\times10^{-7} T^{5/2}$ erg~cm$^{-1}$s$^{-1}$K$^{-1}$ is the heat conductivity parallel to the magnetic field \citep{spit53}. The viscous force is represented as $\vec{F_v}=-\divv \mathbf{\Pi}$, where $\mathbf{\Pi}$ is the viscous stress tensor \citep[see][for details]{ofm22,wan24}.

This study neglects the radiative cooling ($Q_{\rm rad}$) and the empirical heating ($H_{\rm in}$) terms. Instead, an empirical polytropic index value of $\gamma$=1.05 is adopted \citep[e.g.][]{link99}, to reduce the impact of the source terms on the energy equation.

The normalization of the MHD equations results in the dimensionless parameters: the Euler number $E_u ={C_{s0}^2}/{(\gamma V_{A0}^2)}={\beta_0}/{2}$, the Froude number $F_r =  {V_{A0}^2 L_0}/(G M_s)$,  and the Lundquist number $S =\tau_{\rm res}/\tau_A= {L_0 V_{A0}}/{\eta}$, where $\tau_{\rm res}$ is the resistive dissipation time and $\tau_A$ the Alfv\'{e}n transit time. We set $S = 10^4$, implying that the resistivity in the model is much higher than solar resistivity due to numerical limitations,  while the resistive dissipation time is still very long (by 4 orders of magnitude) compared to the Alfv\'{e}n transit time. $L_0$ represents the length scale defined as $L_0=0.1 R_s$ (where $R_s$ is the solar radius). $V_{A0}=B_0/(4\pi\rho_0)^{1/2}$ is the normalizing Alfv\'{e}n speed, $C_{s0}=(\gamma{p_0}/\rho_0)^{1/2}$ is the characteristic sound speed, $G$ denotes the gravitational constant, $M_s$ represents the solar mass, and $\eta$ is the resistivity. The physical parameters used in our numerical model are summarized in Table~\ref{tab:par}.

\begin{table}
\caption{ Physical parameters used in the model.}
\label{tab:par}
\tabcolsep=0pt  \begin{tabular*}{\textwidth}{@{\extracolsep{\fill}}llll@{}}
  \hline\hline
Quantities & Values & Quantities & Values \\
  \hline
Length scale ($L_0$) & 70 Mm & Grav. scale height ($H_0$)  & 61 Mm\\
Magnetic field ($B_0$) & 202 G  & Euler number ($E_u$)  & $1.7\times10^{-4}$\\
Temperature ($T_0$) & 1 MK      & Froude number ($F_r$) & 51.2\\
Number density ($n_0$) & $2\times 10^9$ cm$^{-3}$ & Polytropic index ($\gamma$) & 1.05\\
Alfv\'{e}n speed ($V_{A0}$) & 9852 km~s$^{-1}$ & $\kappa_{\|}(T_0)$ & \makebox[3.1cm][r]{7.8$\times10^{8}$ erg\,{(cm\,s\,K)}$^{-1}$}\\
Alfv\'{e}n time ($\tau_A$) & 7.1 s & $\eta_0(T_0)$ & 0.117 g~{\rm (cm s)}$^{-1}$ \\
        Sound speed ($C_{s0}$) & 131.7 km~s$^{-1}$ & Plasma-$\beta_0$ & $3.4\times 10^{-4}$\\
  \hline
  \end{tabular*}
\end{table}

\subsection{Initial Setup with realistic configuration}
To simulate QFP wave trains observed over a large field of view (FOV) associated with AR 12371, we constructed a 3D coronal AR model initialized within a Cartesian framework using a potential field source surface (PFSS) model based on SDO/HMI photospheric magnetograms.
From the spherical PFSS model, we extracted a cubic volume of extrapolated magnetic fields centered on Carrington coordinates (296$^{\circ}$, 20$^{\circ}$) in longitude and latitude. The simulation domain spans $0.67R_s\times0.67R_s$ at the base, with the center located at a height of 17 Mm above the solar surface. The domain has a total vertical extent of 0.335 $R_s$. The computational grid comprises $612\times612\times306$ high-resolution uniform points. This configuration ensures the preservation of global magnetic connectivity, essential for accurately modeling the observed QFP waves within a large FOV.

The density distribution is initialized based on gravitationally stratified density in polytropic equilibrium, along with the corresponding coronal temperature and base density \citep[see details in][]{ofm22,wan24}. Inspired by EUV observations of fan loops, we construct three open-like loops with footpoints at $z$=0, centered at ($x_0$, $y_0$)=($-$0.6, $-$0.03), ($-$0.3, 0.18), and (0.03, 0.18), with a radius $r_0$=0.15. These modeled loops are defined by tube surfaces that trace magnetic field lines (see Figure~\ref{fig:lpmd}(a)), featuring a peak density ratio $\chi_\rho=\rho_{\rm in}/\rho_{\rm ex}$ and a peak temperature ratio $\chi_T=T_{\rm in}/T_{\rm ex}$. Here, ``in" and ``ex" refer to the interior and exterior of the loop, respectively. For this study, we adopt  $\chi_\rho$=2 and $\chi_T$=1, consistent with the typical physical properties of EUV fan loops \citep{del03, you07, bro11, krish18}. This setup reflects a higher density within the loops compared to the surrounding corona, while maintaining the same temperature distribution. At the loop footpoints, the density across the loop's cross-section is initialized with a sharp Gaussian profile, expressed as:

\begin{equation}
        \rho_{\rm ft}(r)=(\chi_\rho-1)\left(e^{-(r/w)^{2k}}-c_0\right)/(1-c_0)+1.
 \label{eq:sgs}
\end{equation}
Here, $r=\sqrt{(x-x_0)^2+(y-y_0)^2}\le{r_0}$ represents the radial distance from the loop axis in the $xy$-plane at the footpoint. $w=(5/6)r_0$ is the characteristic width of the loop. The sharpness control parameter $k$ is taken as 6. The constant $c_0=e^{-(6/5)^{2k}}$ is selected to smoothly match the loop boundary with the surrounding background corona. Let $\rho_{t=0}^{\rm loop}(r)$ and $T_{t=0}^{\rm loop}(r)$ denote the density and temperature at a point ($x$,$y$,$z$) along a magnetic field line that has the radial distance $r$ at the footpoint to the loop axis. The initial polytropic density and temperature profiles of the loops, satisfying gravitational equilibrium, are given by:
\begin{eqnarray}
        \rho_{t=0}^{\rm loop}&=&\rho_{\rm ft}(r)\left[ 1+\frac{(\gamma -1)}{\gamma H} \left(\frac{1}{10+z-z_{\rm min}} - \frac{1}{10}\right)\right]^{1/(\gamma -1)}=\rho_{\rm ft}(r) \rho_{t=0}^{\rm bg}, \\
  T_{t=0}^{\rm loop}&=& 1+\frac{(\gamma -1)}{\gamma H} \left(\frac{1}{10+z-z_{\rm min}} - \frac{1}{10}\right)= T_{t=0}^{\rm bg},
\end{eqnarray}
where $\rho_{t=0}^{\rm bg}$ and $T_{t=0}^{\rm bg}$ represent the height-dependent background density and temperature, respectively. This indicates that the loop maintains a density contrast $\rho_{\rm ft}(r)$ relative to the surrounding corona throughout its length, while exhibiting no temperature contrast. This loop model establishes a plasma temperature configuration close to the isothermal condition \citep[e.g.,][]{selw11a,selw11b}.

In Equation~\ref{eq:sgs}, the parameter $k$ controls the steepness of the density profile, ranging from $k$=1 for a Gaussian distribution to $k\ge2$ for steeper profiles, approaching a step function for $k\to\infty$. In this study, the loop density profile is set with $k=6$, resulting in a significantly steeper nonuniform layer. This configuration enables the investigation of the effects of transverse loop nonuniformity on wave evolution \citep[e.g.,][]{kolot21}. Figure~\ref{fig:lpmd}(b) depicts the initial state of the coronal loops in the 3D MHD model.

The boundary conditions for the 3D computational domain are set to be open on the four lateral sides, while the bottom and top boundaries are held fixed to enhance numerical stability. Velocity perturbations are applied within a specific region at the bottom boundary to simulate wave driving processes (see Model 1 described in Section~\ref{ssec:wds}).
The simulations are performed in a 3D domain extending in $x$ and $y$ from $-$3.35 to 3.35 and in $z$ from 0 to 3.35, using normalized units. A uniform grid of $612^2\times306$ points is adopted. Experimental control tests demonstrate that using this higher-resolution grid results in substantially lower numerical viscosity than in the $306^3$-grid model. The numerical scheme employs a modified Lax-Wendroff method with a fourth-order stabilization term \citep[e.g.][]{OT02}. Note that the above initial state, including the dense loops, is not an exact stable equilibrium, as the transverse pressure gradient in the loop is not balanced by magnetic pressure in the initial state. However, since the coronal loop is magnetically dominated, characterized by a low-$\beta$ condition, the departure from equilibrium is small. The relaxation process excites strong fast magnetoacoustic waves before we launch periodic perturbations at time $t=10$.

\subsection{Wave Driving Sources}
\label{ssec:wds}

Motivated by the associated QPP analyses in the flare region, we propose two scenarios for the excitation of QFP wave trains. 
The first scenario involves the quasi-separatrix layers (QSLs), which are regions of steep magnetic connectivity gradients that facilitate current sheet formation and magnetic reconnection \citep[e.g.,][]{dem96,dem97}. Theoretical and numerical studies have shown that flare QPPs may be triggered by periodic reconnection at a null point or along a separator \citep[e.g][]{mcl04,door16,sriv25}.  The QSLs can be defined by high-value regions of the squashing factor $Q$, which quantifies the connectivity change of magnetic field lines \citep{tit02,tit07}.  The bottom panels of Figure~\ref{fig:lpmd} show the logarithm $Q$-maps in the $z=0$ and $x=0.53$ planes, calculated from the 3D PFSS AR model using the codes developed by \citet{liur16,zha22}. The Q-maps reveal that the QSLs separate open-like fan loops from closed loops overlying magnetic neutral lines, where the flare ribbons are observed (see Panel~d). The chosen wave-driving region lies at the QSL footprint near the flare ribbons in the lower corona, marked by a smaller blue circle in Panel~(c). 

The second scenario is based on the findings of previous MHD simulations that QFPs can be generated via the interaction of reconnection-produced plasmoids with ambient magnetic fields \citep{yang15, mond24, mond25, liak25}, or of reconnection outflows with the top of flare loops \citep{tak16}.
In this scenario, the wave driver is placed above the magnetic neutral line, where magnetic reconnection occurs in a vertical current sheet, as in the standard flare model. 
This region is indicated by a larger blue circle in Figure~\ref{fig:lpmd}(c).  

 Rather than explicitly simulating the full reconnection dynamics within the QSL or flare current sheet--which would require substantially higher spatial resolution and computational cost--we adopt a spatially localized driver with a harmonic time profile to model the wave excitation, as in earlier QFP modeling studies \citep{ofm11,ofm18,ofm25}. This simplified approach captures the essential physics of wave generation while remaining computationally tractable.
\vspace{3mm}

\noindent {\bf Model 1: Driving by Boundary Perturbations}

Based on the first scenario, and assuming that QPPs occur at the QSL footpoint, we launch periodic velocity pulses at the lower coronal boundary (defined $z$=0) to excite magnetoacoustic waves. The driving source is centered at ($x_0$, $y_0$)= (0, $-$0.5) and defined within a region of $r=\sqrt{(x-x_0)^2+(y-y_0)^2}\le w=0.15$, with the following form:
 \begin{equation}
         V_y(x,y,t\ge{t_1})  = V_0\left[\frac{1}{2}\left(1 - \cos \frac{2 \pi (t-t_1)}{P_0}\right) \right] \exp\left[ -\left( \frac{r}{w} \right)^2 \right],
         \label{eq:wdr1}
\end{equation}
where $V_0$ is the amplitude, $P_0=10$ the period, and $w$ the source region's half-width. 

The perturbation starts at $t_1=10$, to allow the initial state to relax. The flow pulses are directed in the $y$-direction, reflecting the observed QFP wave propagation predominantly toward the north and northeast. The driver amplitude $V_0=0.01$ is chosen such that the simulated QFP waves produce relative density perturbations comparable to those observed.  Varying $V_0$ by factors of a few above or below this value produces waves with qualitatively similar propagation characteristics.

\vspace{5mm}

\noindent  {\bf Model 2: Driving by Periodic Force}

In the second scenario, we model the wave excitation using a spatially localized periodic force. This imposed driver is a reduced representation of quasi-periodic disturbances expected from bursty magnetic reconnection and plasmoid/outflow interactions above the neutral line. The force follows a Gaussian spatial amplitude distribution that approximates the localized energy or momentum deposition, and is centered at ($x_0$,$y_0$,$z_0$) = (0.7, $-$0.6, 0.3), just above the flare-associated magnetic neutral line (see Figure~\ref{fig:bimg}B and Figure~\ref{fig:lpmd}c).
This forcing is implemented by adding the following term to the right-hand side of the moment equation (Equation~\ref{eq:mm}):
 \begin{equation}
         F_z(x,y,z,t\ge{t_1})  = - f_0 \sin \frac{2 \pi (t-t_1)}{P_0} \exp\left[ -\left( \frac{r}{w} \right)^2 \right], 
         \label{eq:wdr2}
 \end{equation}
where $r=\sqrt{(x-x_0)^2+(y-y_0)^2+(z-z_0)^2}$. $f_0$ is the force amplitude, $P_0=10$ is the driving period, $w=0.15$ is the characteristic scale of the source, and $t_1=10$ represents the onset time of the perturbation. Because a larger external forcing (i.e., a larger value of $f_0$) would require smaller time steps and higher spatial resolution to accurately resolve rapid variations near the injected wave energy flux, leading to prohibitively long computation times, we adopt $f_0=0.007 $ in this study. As a consequence, the resulting QFP waves exhibit relative density perturbations that are several times weaker than those observed (see Section~\ref{ssc:md2}). One way to address this issue is to adopt a non-uniform grid; however, such an improvement is beyond the scope of the current computational capabilities of this code.

\section{Numerical Results} 
\label{sec:result}

\subsection{Evolution of Physical Parameters for Model 1}
\label{ssc:epp}

\begin{figure*}
\gridline{\fig{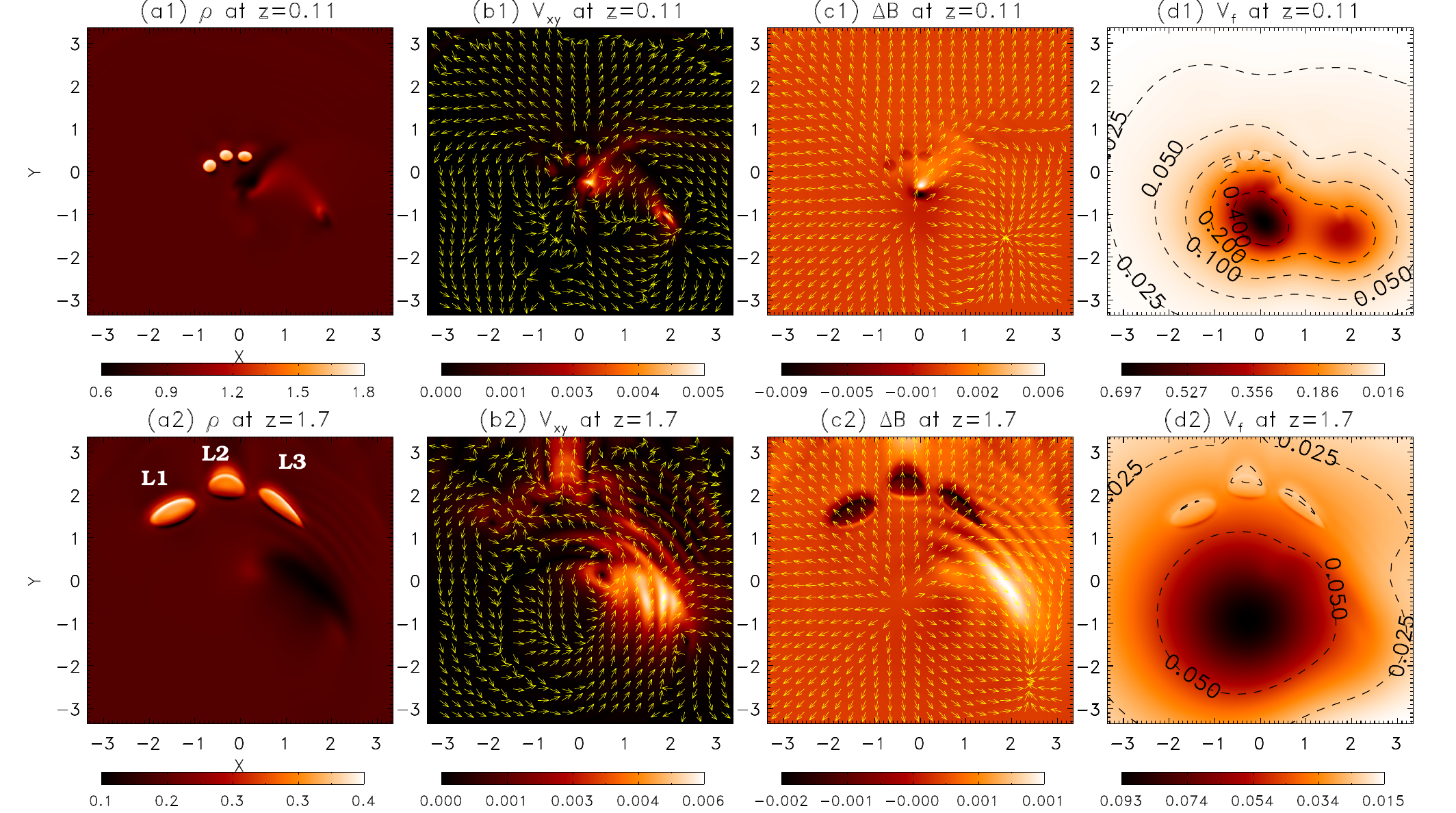}{0.9\textwidth}{}}
\gridline{\fig{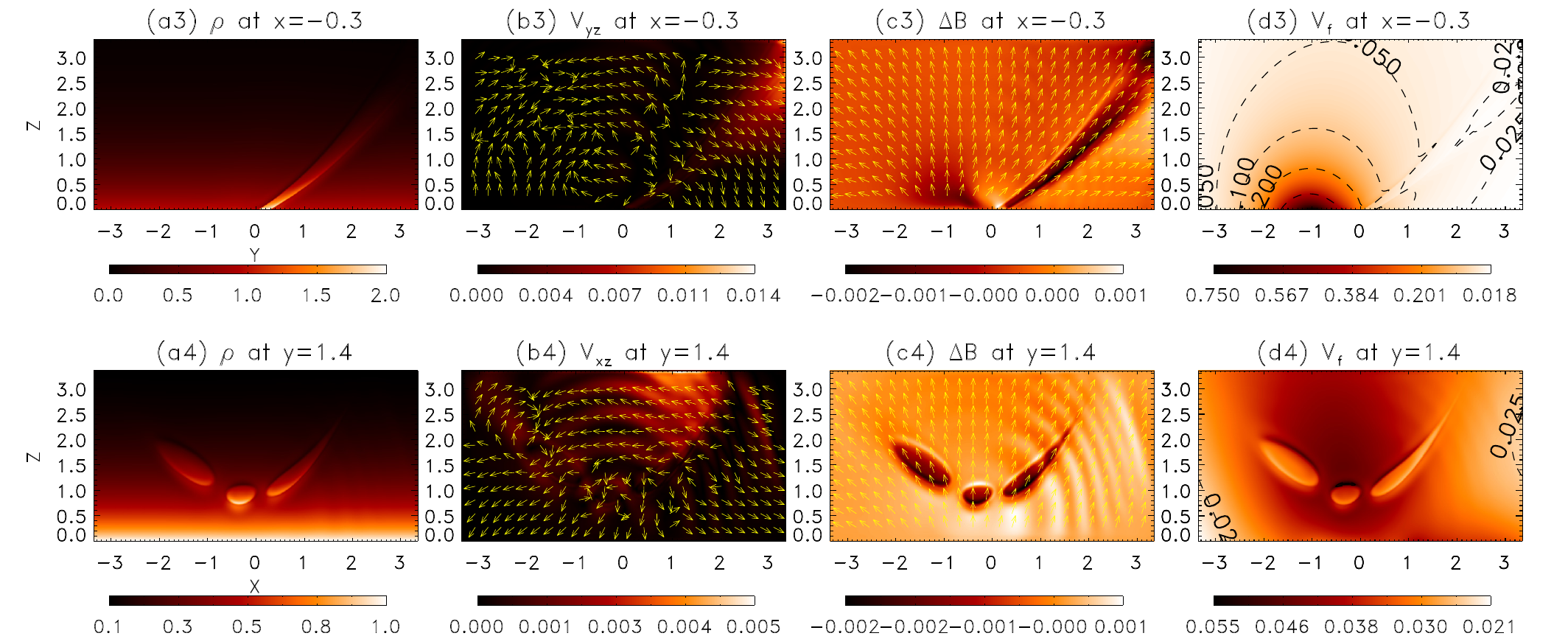}{0.9\textwidth}{}}
\caption{Results of 3D MHD simulations of QFP waves for Model~1. \textit{Top two rows:} Snapshots show distributions of physical parameters on horizontal cross-sections at two different heights. Panels (a1)-(d1) show density, horizontal velocity, magnetic perturbation, and fast magnetoacoustic speed, respectively, in the $xy$-plane at $z=0.11$ and $t=100\tau_A$. Panels (a2)-(d2) shows same parameters as in (a1)-(d1) but at $z=1.7$. In (a2), the three dense loops are labeled $L1$, $L2$, and $L3$.  \textit{Bottom two rows:} Snapshots of physical parameter distributions on two perpendicular vertical cross-sections are shown. Panels (a3)-(d3) display density, velocity component, magnetic perturbation, and fast magnetoacoustic speed, respectively in the $yz$-plane at $x=-0.3$ and $t=100\tau_A$. Panels (a4)-(d4) shows same parameters as in (a3)-(d3) but in the $xz$-plane at $y=1.4$. In (b1)-(b4), the color scale represents the velocity magnitude, and arrows indicate the velocity direction. In (c1)-(c4), the color scale represents the magnitude of the perturbed magnetic field ($\Delta B = B(t)-B(0)$), and arrows show the direction of the magnetic field components in the plane at time $t$. In (d1)-(d4), darker regions indicate higher values, and contours correspond to $V_f = 0.025$, 0.05, 0.1, 0.2, and 0.4.      
\label{fig:mpar}}
\end{figure*}

We first present the simulation results for Model 1, in which QFP wave trains are driven by perturbations at the bottom boundary. Top two rows of Figure~\ref{fig:mpar} shows the distributions of physical parameters at two layers ($z = 0.11$ and 1.7) at time $t = 100\tau_A$: density (panels a1 and a2), velocity (panels b1 and b2), magnetic perturbation (panels c1 and c2), and fast magnetoacoustic speed (panels d1 and d2). The fast magnetoacoustic speed is calculated as $V_f = \sqrt{V_A^2 + C_s^2}$, representing the maximum phase speed when waves propagate perpendicular to the local magnetic field. The three bright patches in panels (a1) and (a2) mark the cross-sections of the modeled dense coronal loops ($L1$, $L2$, and $L3$, labeled in panel a2). Wave perturbations propagating northwestward are clearly visible in panels (a2)–(c2) at the higher height. The fast-wave speeds at the lower height (panel d1) are highly anisotropic and decrease rapidly away from the AR center, whereas the speeds at the higher height (panel d2) become more isotropic and decrease more gradually. Panel (d2) further indicates that the fast-wave speeds inside the dense loops are slightly lower than those in the surrounding plasma.

Bottom two rows of Figure~\ref{fig:mpar} present the distributions of density, velocity, magnetic perturbation, and fast magnetoacoustic speed at $t = 100 \tau_A$, in cross-sections at $x = -0.3$ (panels a3–d3) and $y = 1.4$ (panels a4–d4). Density stratification with height is evident both in the background and within the loops (panels a3 and a4). Panels (c3) and (c4) show that the magnetic field inside the dense loops is slightly weaker than outside, a result of magnetic-pressure adjustment following the initial density enhancement of the loops. Rebalancing of total pressure between the loop interior and exterior leads to this modest reduction in magnetic pressure inside. Upward and westward propagation of fast waves is clearly visible in panels (b4) and (c4). The $x$-cut across the AR core reveals that the fast-wave speed decreases rapidly with height (panel d3), whereas the $y$-cut across the AR edge shows a relatively uniform distribution, with slightly weaker speeds inside the loops (panel d4).

\begin{figure*}
\gridline{\fig{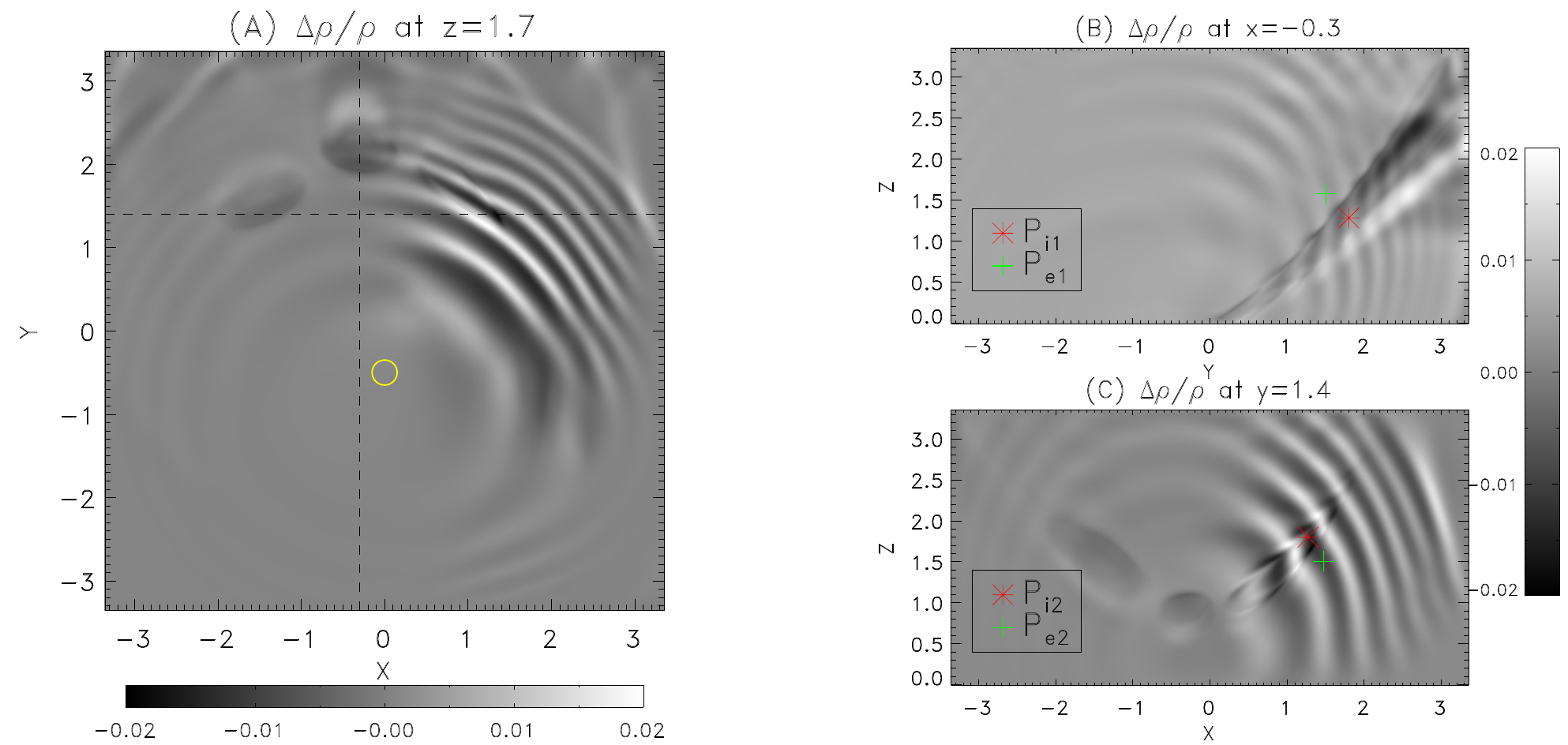}{0.7\textwidth}{}}
\gridline{\fig{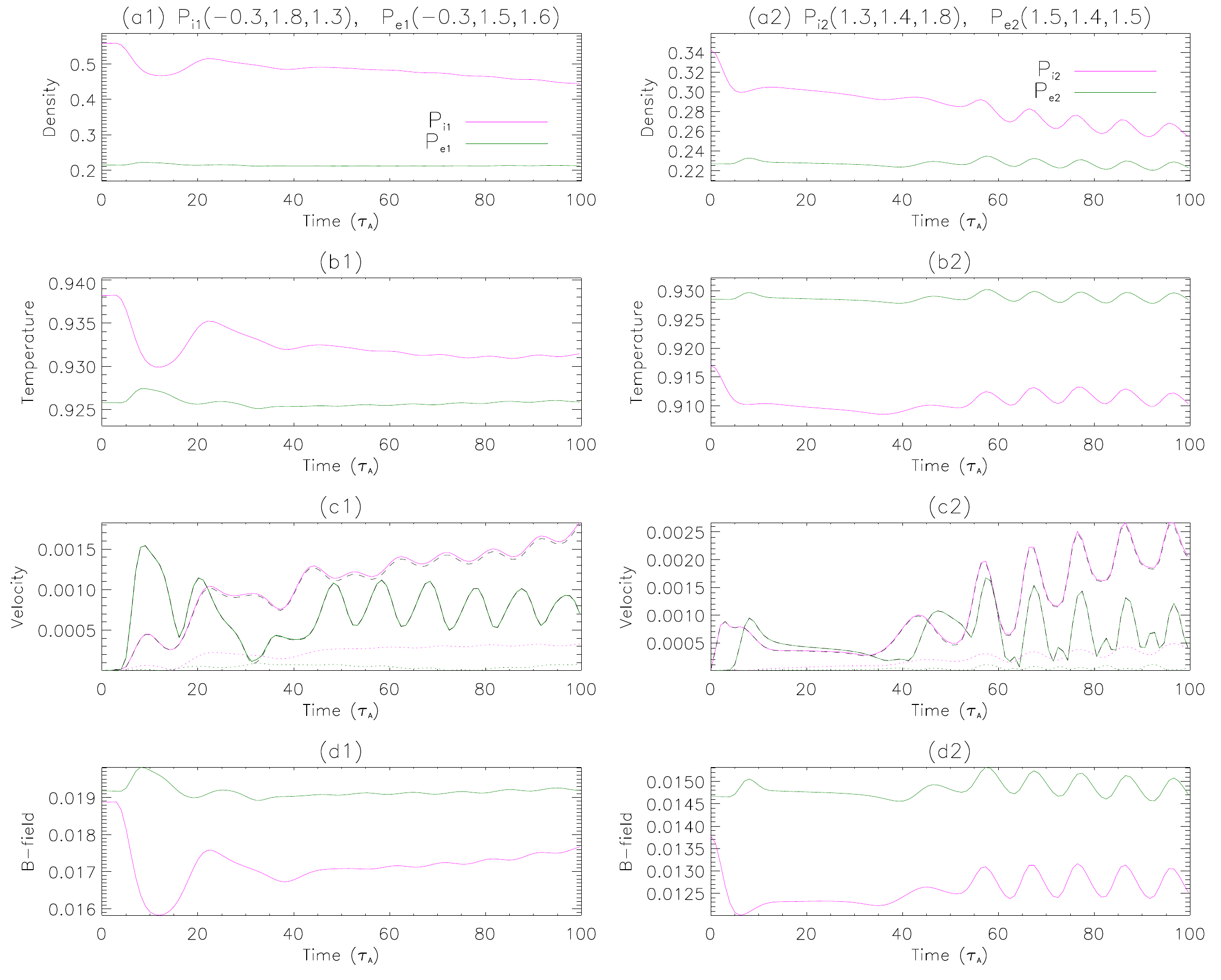}{0.7\textwidth}{}}
\caption{\textit{Top panels:} Snapshots of density distributions in relative running difference ($\Delta \rho / \rho = [\rho(100\tau_A)-\rho(99\tau_A)] / \rho(100\tau_A)$) at three cross-sectional planes: (A) $xy$-plane at $z=1.7$, (B) $yz$-plane at $x=-0.3$, and (C) $xz$-plane at $y=1.4$. Vertical and horizontal dashed lines indicate the cut positions for density maps shown in (B) and (C), respectively. A small circle marks the location of the velocity driver in the $xy$-plane.  In (B) and (C), two positions (marked by an asterisk and a plus sign, located inside and outside the dense loops, respectively) are selected to illustrate the time profiles of variables. The accelerated animation of panels (A)-(C) presents the simulation over the time interval $t=(1-100)\tau_A$, corresponding to a duration of 710 s. Panels (a1)-(d1) show time profiles of density, temperature, total velocity, and total magnetic field, respectively, at positions $P_{i1}$ and $P_{e1}$. In panel (c1), the dashed line indicates the velocity component perpendicular to the magnetic field (nearly overlapping with the solid line), while the dotted line denotes the velocity component parallel to the magnetic field. Panels (a2)-(d2) show same as in (a1)-(d1), but for positions $P_{i2}$ and $P_{e2}$.  (An animation of this figure is available in the online article.)      \label{fig:ncut}}
\end{figure*}

To illustrate wave propagation inside and outside the loops, Figure~\ref{fig:ncut}(A)-(C) show relative running-difference maps of density at three cross-sectional planes. These maps are defined as $\Delta \rho / \rho = [\rho(100\tau_A) - \rho(99\tau_A)] / \rho(100\tau_A)$). Animations of panels (A)-(C), showing the wave evolution from $t = 0$ to $100 \tau_A$, are available in the online journal. This indicates that the waves propagate with clear directionality and weak damping. The wave fronts become distorted as they cross or pass through the loops due to the sharp change in fast-mode speed at the loop boundaries. Overall, however, the dense loops exert only a minor influence on the propagation of the excited waves.

To compare wave behavior inside and outside the loops, we plot time profiles of physical parameters at two selected locations marked in Figure~\ref{fig:ncut}(B) and (C). Points $P_{i1}$ and $P_{e1}$ lie inside and outside loop L2, respectively, while $P_{i2}$ and $P_{e2}$ are located inside and outside loop L3. Figure~\ref{fig:ncut}(a1)–(d1) shows the evolution of density, temperature, total velocity, and total magnetic field for $P_{i1}$ and $P_{e1}$, and Figure~\ref{fig:ncut}(a2)–(d2) shows the corresponding profiles for $P_{i2}$ and $P_{e2}$. In both cases, the variations for $t < 30 \tau_A$ are due to disturbances from the relaxation of the initially inserted loops. For $P_{i1}$ and $P_{e1}$, the wave signatures are weak, evident only in velocity, with smaller amplitudes inside the loop. In contrast, for $P_{i2}$ and $P_{e2}$, strong wave signals appear in all parameters, with amplitudes slightly larger inside than outside, consistent with wave propagation primarily in the northwestern direction (see Figure~\ref{fig:ncut}a). From the oscillations at $P_{i2}$ and $P_{e2}$, we measure relative amplitudes of $\Delta{\rho}/\rho=(3.0\pm0.6)\%$ and $(2.0\pm0.2)\%$ in density, $\Delta{T}/T=(0.15\pm0.03)\%$ and $(0.10\pm0.01)\%$ in temperature, and $\Delta{B}/B=(2.7\pm0.3)\%$ and $(1.8\pm0.1)\%$ in magnetic field, respectively, while the velocity amplitudes are  $5.4\pm1.7$ and $6.2\pm1.0$ km s$^{-1}$, respectively.

Figure~\ref{fig:ncut}(a2)–(d2) shows that the density, temperature, and magnetic field oscillate nearly in phase, consistent with fast-mode waves in which the magnetic and gas pressures vary together. Figure~\ref{fig:ncut}(c1) and (c2) show that the velocity perturbations perpendicular to the magnetic field are significantly larger than those parallel to it, indicating that the waves propagate predominantly across the magnetic field at the analyzed locations. The measured amplitudes of the perturbations further reveal the relationship, $\Delta{\rho}/\rho\approx\Delta{B}/B$, which provides additional support for this conclusion.  

Furthermore, a control run of Model 1 with thermal conduction and compressive viscosity switched off yields results for the physical parameters nearly identical to those obtained using classical coefficient values (not shown). This suggests that, under the present conditions, thermal conduction and compressive viscosity play a negligible role in wave dissipation.

\subsection{Evolution of Synthetic Emission Measure for Model 1}
\label{ssc:eem}
\begin{figure*}
\plotone{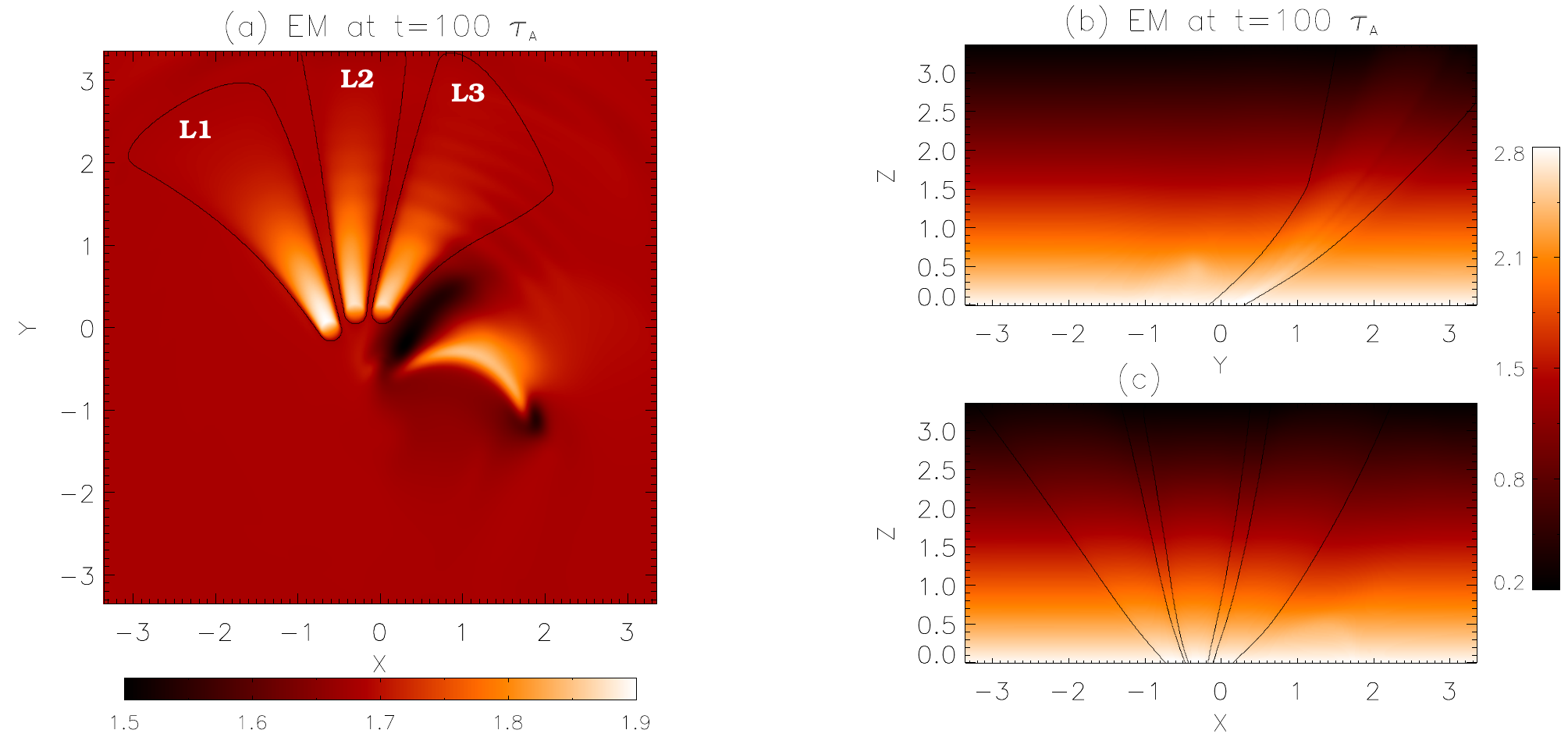}
\caption{Snapshots of emission measure (EM) distributions viewed along three orthogonal directions at $t=100\tau_A$. (a) EM in the $xy$-plane (${\rm EM}_{xy}=\int \rho^2\,dz$). (b) EM in the $yz$-plane (${\rm EM}_{yz}=\int \rho^2\,dx$). (c) EM in the $xz$-plane (${\rm EM}_{xz}=\int \rho^2\,dy$). The black contours delineate the fan loop regions, defined as areas where the initial EM exceeds the background level by 0.1\%.        
\label{fig:emd1}}
\end{figure*}

To compare the modeled results with observations, we compute the emission measure (EM) from the simulated 3D density distribution as $EM = \int_{\mathrm{LOS}} \rho^2(x,y,z) dl$, where $l$ is the distance along the line of sight (LOS). Figure~\ref{fig:emd1} shows the EM distributions in three projected planes at time $t = 100 \tau_A$. In panel (a), three fanlike coronal loops (labeled $L1$, $L2$, and $L3$) together with the excited waves are clearly visible, corresponding to the on-disk view. Panel (a) also reveals a closed dense loop in the AR core, produced by periodic flow pulses near its eastern footpoint. In contrast, the modeled loops and waves are hardly discernible in panels (b) and (c), which correspond to limb views, because of strong background emission.

For consistency in analysis, we define the ``observed" loops as structures whose initial EM exceeds the background level by more than $0.1\%$. The background EM is defined as follows: for ${\rm EM}_{xy}$, ${\rm EM}_{\mathrm{bg}} = {\rm EM}(x, y_{\min})$; for ${\rm EM}_{yz}$, ${\rm EM}_{\mathrm{bg}} = {\rm EM}(y_{\min}, z)$; and for ${\rm EM}_{xz}$, ${\rm EM}_{\mathrm{bg}} = {\rm EM}(x_{\max}, z)$.

\begin{figure*}
\gridline{\fig{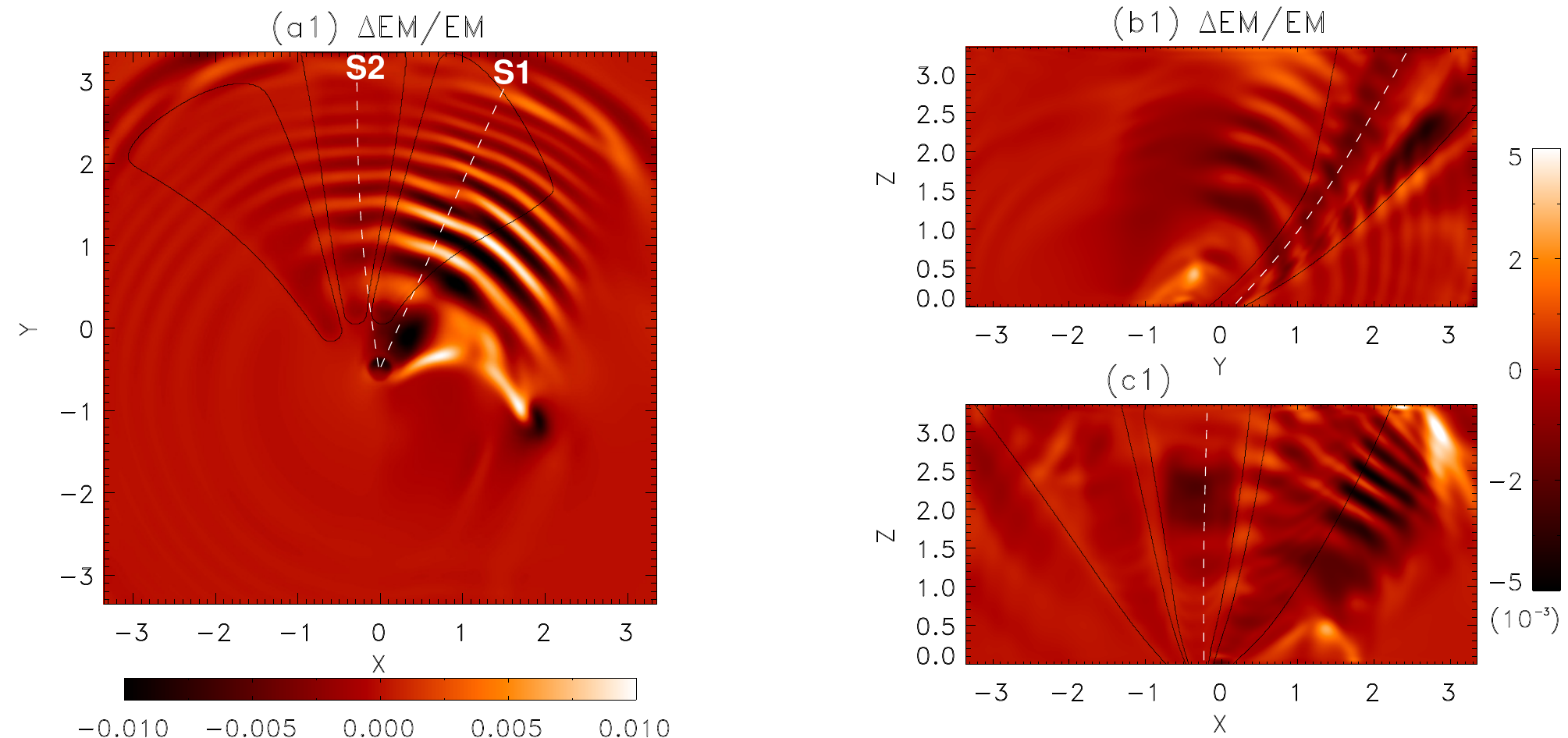}{0.9\textwidth}{}}
\gridline{\fig{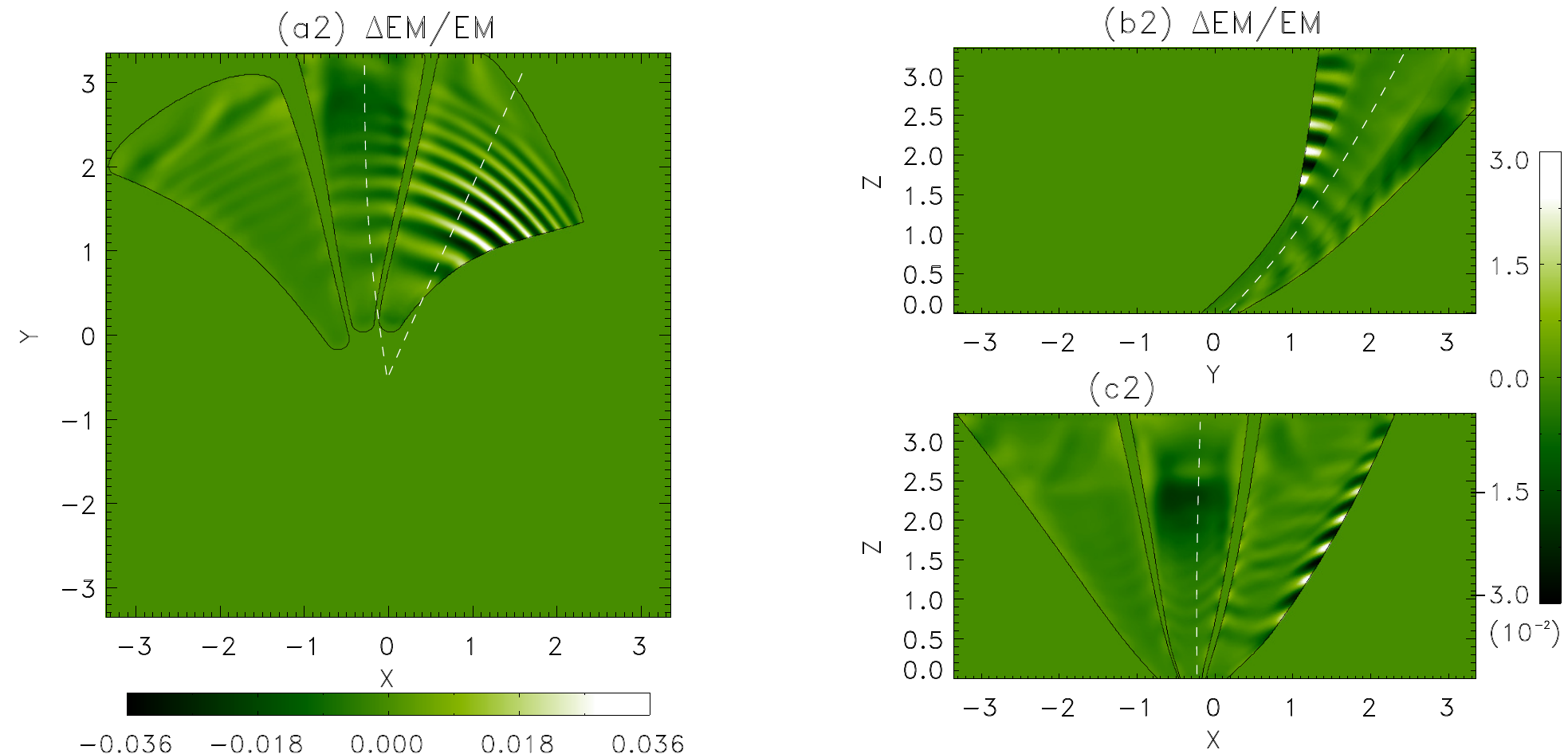}{0.9\textwidth}{}}
\caption{Results of 3D MHD simulations of the QFP waves for Model~1. \textit{Top panels:} Snapshots of emission measure distributions viewed along three orthogonal directions at $t=100\tau_A$, shown in relative running difference form ($\Delta{\rm EM}/{\rm EM}=({\rm EM}(100\tau_A)-{\rm EM}(99\tau_A))/{\rm EM}(100\tau_A)$): (a1) in the $xy$-plane, (b1) in the $yz$-plane, and (c1) in the $xz$-plane. The black contours indicate the same fan loop regions as described in Figure~\ref{fig:emd1}. The white dashed lines mark the slice positions used to produce the time-distance maps in Figure~\ref{fig:tds1}. \textit{Bottom panels:}  Same as the top panels, but showing the EM distributions contributed exclusively by the fan loop plasma. The black contours outline the fan loop regions where the initial EM exceeds zero.  The accelerated animation presents the simulations for Model 1 in two cases: one including the coronal background (Panels a1--c1) and one including only the fan-loop contribution (Panels a2--c2), over the time interval $t=(1-100)\tau_A$, corresponding to a duration of 710 s.  (An animation of this figure is available in the online article.)   \label{fig:rdf1}}
\end{figure*}

Figure~\ref{fig:rdf1}(a1)-(c1) presents snapshots of EM distributions in the relative running difference, projected on the $xy$-, $yz$-, and $xz$-planes at $t=100\tau_A$. The waves are seen to propagate both inside and outside the coronal loops, mainly in the northwestern direction in the on-disk view (panel a1), and upward into the higher corona in the limb views (panels b1 and c1). Animations showing the wave evolution from $t=0$ to $100\tau_A$ are available in the online journal.

The waves appear more prominent in the top view (panel a1) than in the side views. This can be explained by the fact that wave fronts at lower heights tend to be nearly vertical; when viewed along the side, e.g., the $y$-direction, density perturbations in peaks and valleys tend to cancel each other in the LOS integration (see the weak wave signals inside the loops in panel c1). A similar effect likely accounts for the weak signals in the $x$-direction view (panel b1), where the LOS overlaps with three loops. Moreover, Figure~\ref{fig:rdf1}(b1) shows that the upward-propagating waves have noticeably longer wavelengths than the northward-propagating ones, implying that the former propagate faster than the latter--consistent with the distribution of fast-mode speeds (Figure~\ref{fig:mpar}d3). 

It is well known that the SDO/AIA EUV channels respond to plasma at different characteristic temperatures, since each channel is dominated by emission lines of specific ions. For instance, the 171~\AA\ channel is dominated by Fe\,{\sc ix}, with peak sensitivity at $\sim$0.6~MK; the 193~\AA\ channel is dominated by Fe\,{\sc xii}, peaking at $\sim$1.5~MK; and the 211~\AA\ channel is dominated by Fe\,{\sc xiv}, peaking at $\sim$2~MK. Consequently, these channels exhibit different response sensitivities to coronal loops and background emission. For example, fan loops typically have electron temperatures of $T_e=0.6$--1~MK, the quiet-Sun (QS) corona has $T_e=1$--1.5~MK, and AR corona has $T_e=2$--4~MK \citep{asch04}. As a result, the fan loops are more distinctly visible in 171 \AA\ than in 193 \AA, since the 193 \AA\ channel is more sensitive to the diffusive coronal emission from non-loop plasma.  

For clarity, hereafter we refer to {\it the coronal background} as the diffuse coronal plasma emission along the LOS that is {\it not} part of the specific structure under study (e.g., loops), including contributions from both foreground and background plasma.

To investigate how the coronal background influences the observed wave signatures as the waves propagate through different coronal structures, we calculated the EM distributions contributed exclusively by the fan loop plasma, ${\rm EM}_{\rm loop}=\int_{\rm LOS}{\rho_{\rm loop}(x,y,z)^2}dl$, where $\rho_{\mathrm{loop}}(x,y,z)$ represents the density distribution inside the fan loops, with volumes assumed to remain the same as initially. The ${\rm EM}_{\rm loop}$ corresponds to a case similar to AIA 171 \AA\ observations. Figure~\ref{fig:rdf1}(a2)-(c2) shows the EM distributions of the modeled fan loops in relative running difference, projected onto the three orthogonal planes at $t=100 \tau_A$. The wave features observed within the modeled loops differ markedly between the cases with and without the coronal background. For example, in loop $L3$ viewed in the $xy$-plane, the apparent wavelengths are shorter when the coronal background is excluded than when it is included. Likewise, the wave features viewed from the sides differ substantially between the two cases (compare panels b2 and c2 with b1 and c1, respectively). The case containing only coronal loops fails to capture the prominent upward-propagating wave signals near the loops. These results indicate that the response sensitivity to coronal background emission significantly affects the observed wave features in EM for certain channels.

\begin{figure*}
\gridline{\fig{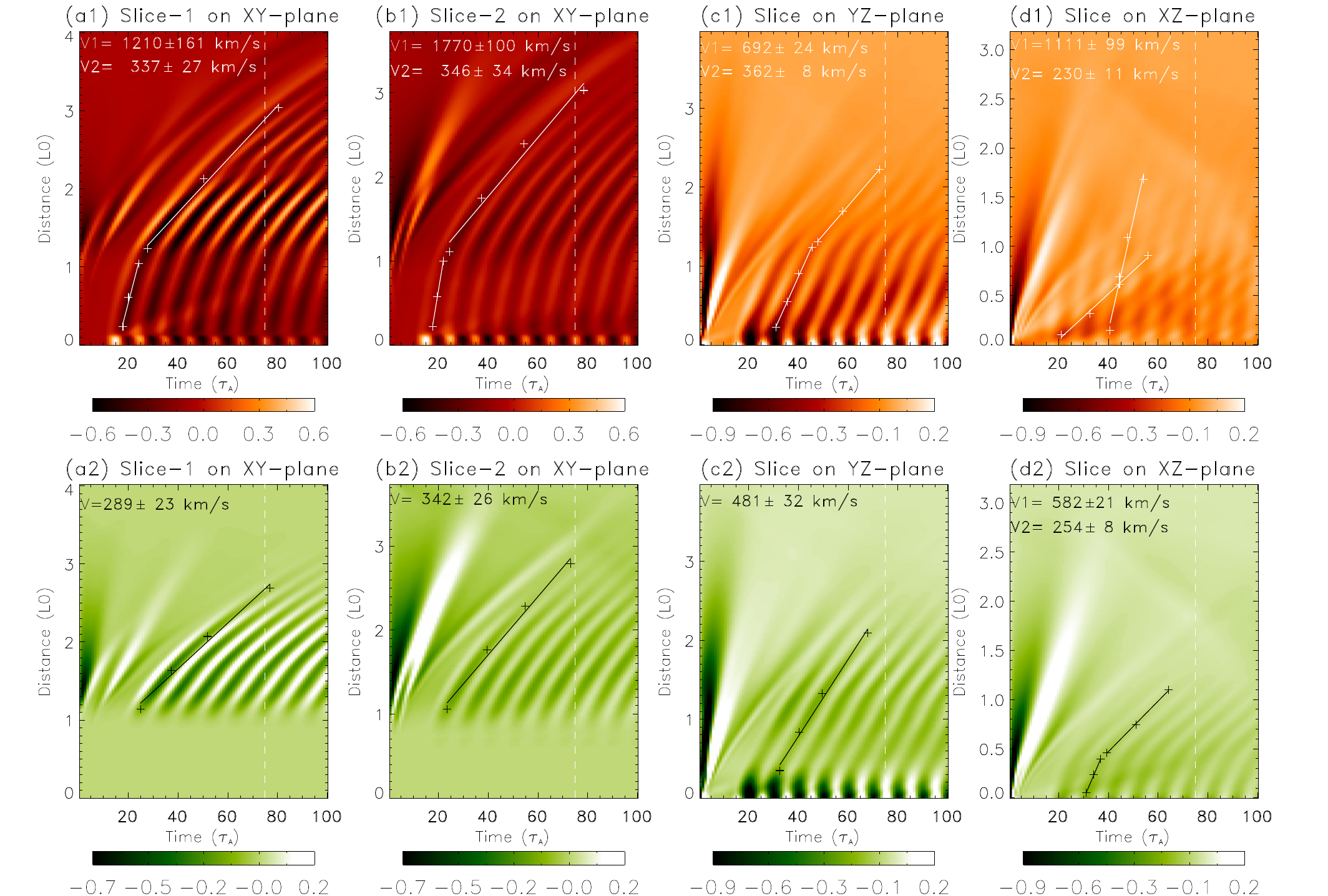}{0.8\textwidth}{}}
\gridline{\fig{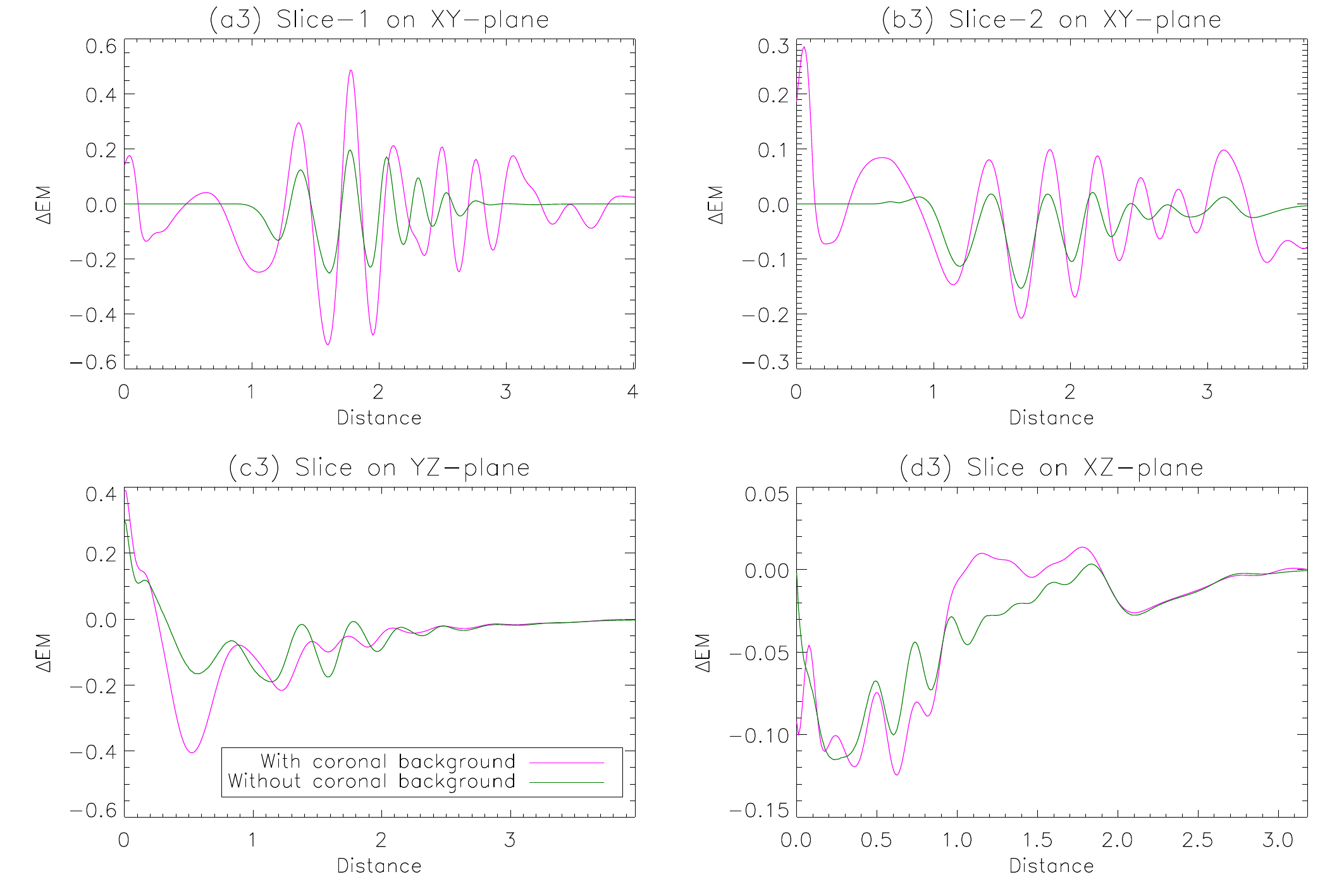}{0.8\textwidth}{}} 
\caption{Panels (a1)-(d1): Time-distance maps of the emission measure in running difference ($\Delta {\rm EM}$) along the slices shown in Figure~\ref{fig:rdf1}. Slice~1 and Slice~2, used to generate panels (a1) and (b1), respectively, are indicated as S1 and S2 in Figure~\ref{fig:rdf1}(a1). Panels (a2)-(d2) show same as in (a1)-(d1), but for the emission measure distributions contributed exclusively by the fan loop plasma. The crosses and linear fits (solid lines) are used to measure the wave propagation speeds. Panels (a3)-(d3): spatial profiles of the emission measure in running difference at $t=75\tau_A$ along the vertical cuts shown in the above panels. Panels (a3) and (b3) show $\Delta{\rm EM}$ along Slice~1 and Slice~2 in the $xy$-plane, respectively; panel (c3) shows $\Delta{\rm EM}$ along the slice in the $yz$-plane, and panel (d3) along the slice in the $xz$-plane. The purple line represents the case where the coronal background is included in the EM calculation, while the green line represents the case where it is excluded.        \label{fig:tds1}}
\end{figure*}

Top two rows of Figure~\ref{fig:tds1} show the time-distance maps of EM along four slices (averaged over a narrow width of 0.055) on three projected planes, as shown in Figures~\ref{fig:rdf1}. Slice-1 and Slice-2, labeled with S1 and S2 on the $xy$-plane, are selected starting from the driving source and primarily following loop-L3 and loop-L2, respectively. Additional slices on the $yz$- and $xz$-planes are also taken along the simulated loops. 

The wave characteristics differ markedly between the case that includes the coronal background in the EM calculation (panels a1–d1) and the case that considers only the loop emission (panels a2–d2). Panels (a1) and (b1) show that the waves propagate from distance $s$ = 0 to $\sim1$ at a very high speed, completing the travel in less than one wave period. The wave speeds over this distance are measured to be $1210\pm161$ km~s$^{-1}$ and $1770\pm100$ km~s$^{-1}$ from the time-distance plots of Slice-1 and Slice-2, respectively. 

In contrast, for distances $s \gtrsim 1$, the wave propagation speeds decrease, with values estimated at $337 \pm 27$ km s$^{-1}$ and $346 \pm 34$ km s$^{-1}$ from Slice-1 and Slice-2, respectively. Panel (c1) shows that the waves propagate more rapidly over the range $s = 0$ to $\sim 1.2$, with an estimated speed of $692 \pm 24$ km s$^{-1}$, compared to $362 \pm 8$ km s$^{-1}$ for $s > 1.2$. Panel (d1) reveals the superposition of two wave components traveling at different speeds: a faster component at $1111 \pm 99$ km s$^{-1}$ and a slower one at $230 \pm 11$ km s$^{-1}$. 

Panels (a2) and (b2) indicate that the waves appear to originate at $s \approx 1$, since the background emission is not included in this case. This behavior is consistent with the AIA 171 \AA\ observations (see Figure~\ref{fig:tdismap}), as this channel is not sensitive to hot plasma emission near the AR core. The wave propagation speeds measured from the time-distance plots of panels (a2) and (b2) are $289 \pm 23$ km s$^{-1}$ and $342 \pm 26$ km s$^{-1}$, respectively. In panel (c2), the propagation speed is about $481 \pm 32$ km s$^{-1}$. In panel (d2), the waves travel faster over $s = 0$ to $\sim 0.5$ at $582 \pm 21$ km s$^{-1}$, but slow down for $s > 0.5$ to $254 \pm 8$ km s$^{-1}$.

Careful comparison between the cases with (panels a1–d1) and without (panels a2–d2) the coronal background reveals clear differences in wave behavior. In the former case, the superposition of propagating waves inside and outside the loops modifies the apparent wave amplitudes, propagation speeds, and phases relative to the latter. This suggests that variations in density and temperature among different coronal structures (e.g., AR loops and the quiet Sun), combined with the temperature selectivity of instrument response functions, may give rise to pseudo-wave characteristics. 

In addition, panels (b1)–(d1) and (b2)–(d2) both display an initial broad front propagating at a speed significantly higher than that of the subsequent waves. The propagation speeds measured from panels (b1)–(d1) are $645 \pm 13$, $915 \pm 30$, and $691 \pm 17$ km s$^{-1}$, respectively. This front, generated by the relaxation of the initially inserted dense loops, exhibits properties consistent with fast-mode shocks.

Bottom two rows of Figure~\ref{fig:tds1} compare the spatial profiles of EM along the four slices analyzed above for time $t=75 \tau_A$ between the two cases with and without coronal background. Quantitatively, it indicates different wave behaviors in the two cases. Panels (a3) and (b3) clearly show that the waves propagate at a markedly high speed from the driving source to $s \approx 1$, characterized by a very long wavelength over this range. According to wave flux conservation, the higher phase speed corresponds to a relatively lower amplitude compared to that of the subsequent, slower waves. This property may also explain why flare-excited waves traveling near the AR core are difficult to detect in actual observations.

Figure~\ref{fig:tds1} (a3) and (b3) also shows that, when viewed from above, the waves in the case including the coronal background exhibit larger amplitudes, propagate faster, and decay more slowly than in the case considering only loop emissions. This behavior can be explained by the fact that the fast-mode speed is relatively lower inside the loop than outside, and that waves traveling inside the loop change their propagation direction depending on the loop geometry. When the waves propagate predominantly upward at a certain height, the corresponding on-disk observations would show them as apparently damped. Consequently, for waves traveling over the range $s \gtrsim 2$ in the former case, the apparent wave signatures are likely dominated by the component propagating outside the loops.

Panels (c3) and (d3) show that, when viewed from the side, the waves exhibit relatively smaller amplitudes and appear to damp more quickly than when viewed from above. This is consistent with the fact that the waves propagate predominantly in the horizontal direction rather than upward in this simulated example.

\subsection{Evolution of Synthetic Emission Measure for Model 2}
\label{ssc:md2}

\begin{figure*}
\gridline{\fig{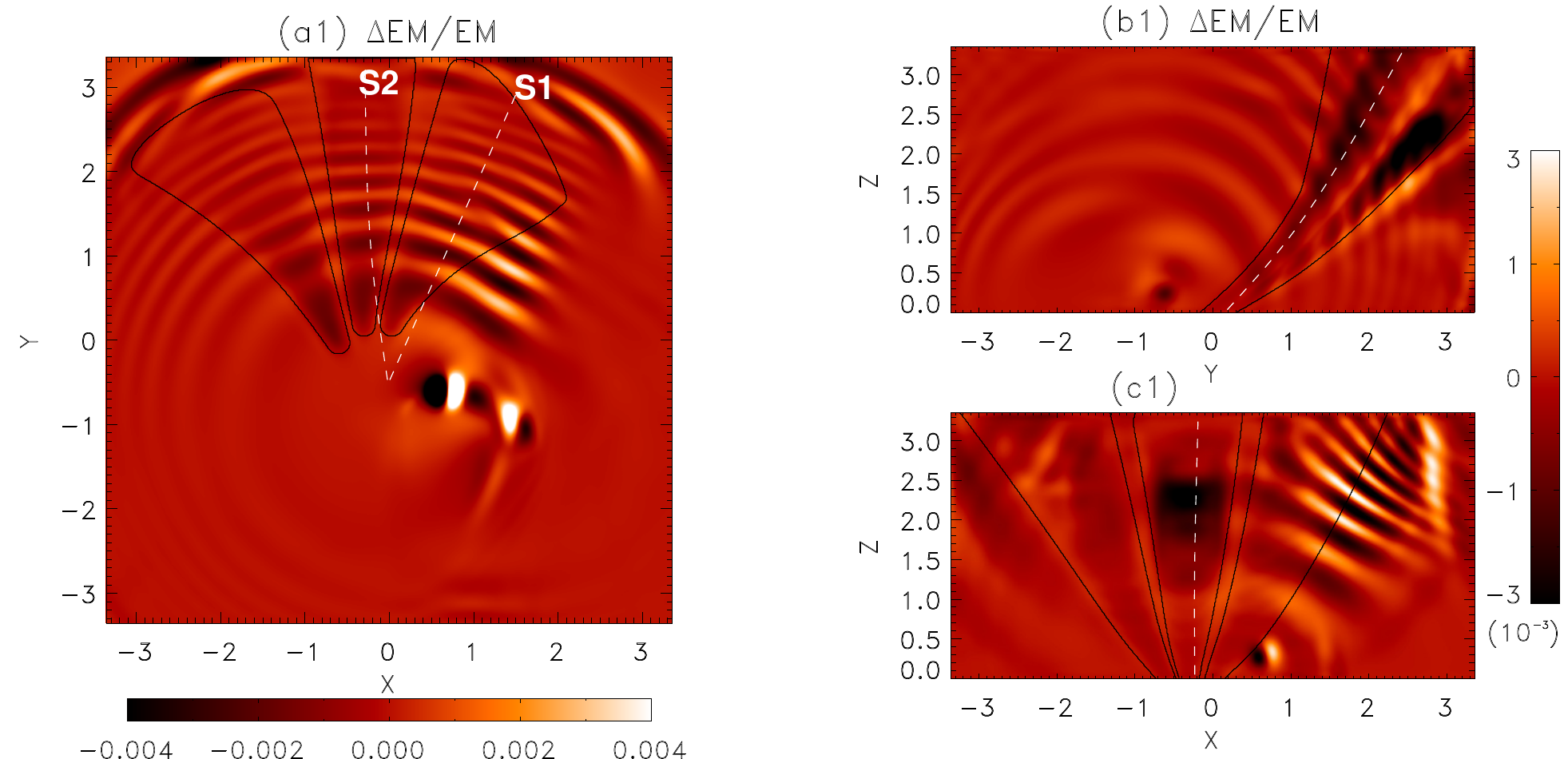}{0.9\textwidth}{}}
\gridline{\fig{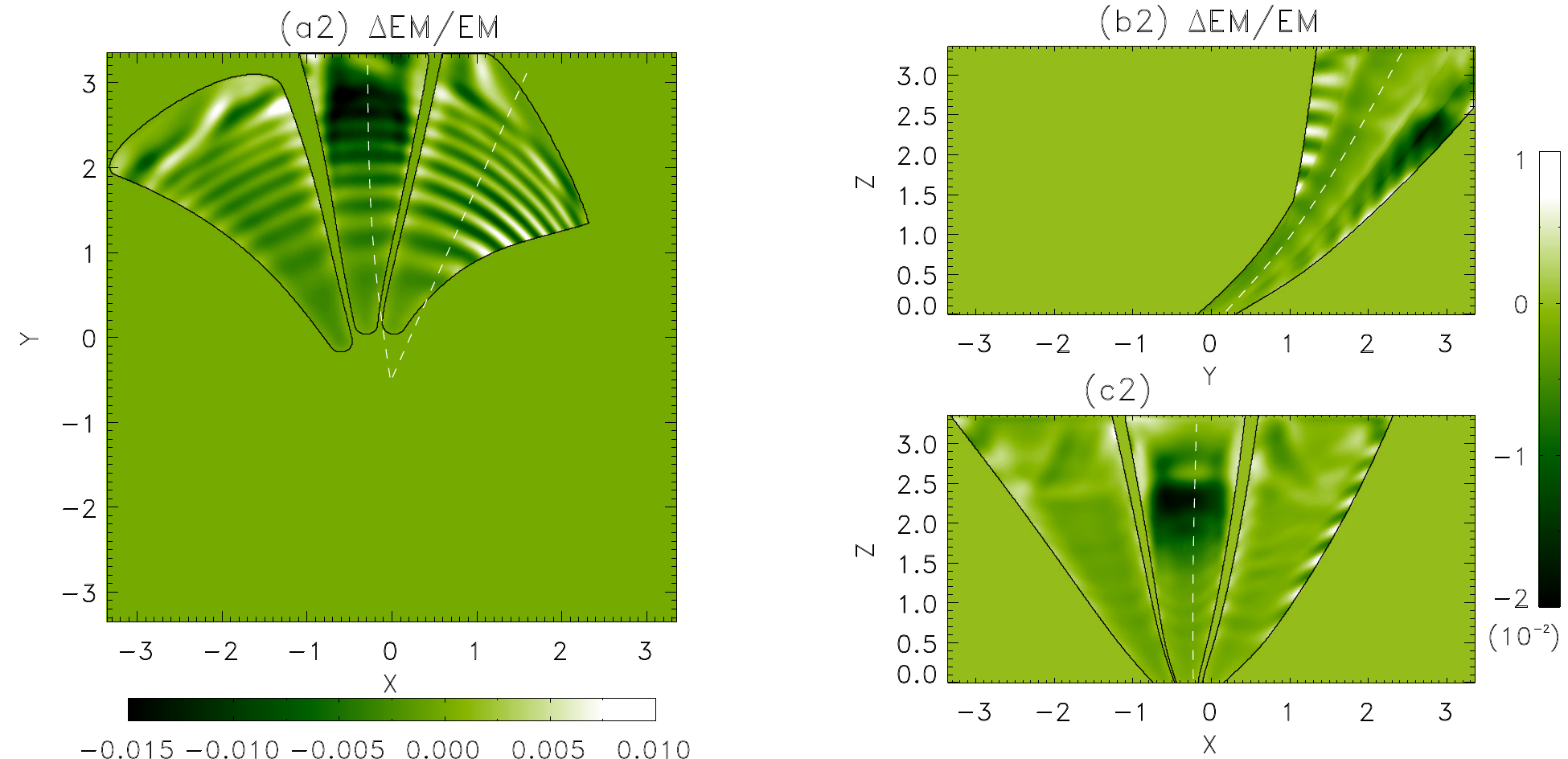}{0.9\textwidth}{}}
\caption{Results of 3D MHD simulations of the QFP waves for Model~2. \textit{Top panels:} Snapshots of the EM distributions viewed along three orthogonal directions at $t=100\tau_A$, shown in running difference ($\Delta{\rm EM}/{\rm EM}$): (a1) in the $xy$-plane, (b1) in the $yz$-plane, and (c1) in the $xz$-plane. The black contours indicate the same features as described in Figure~\ref{fig:emd1}. The white dashed lines mark the slice positions used to create the time-distance maps shown in Figure~\ref{fig:tds2}.  \textit{Bottom panels:}  Same as the top panels, but for the EM distributions contributed exclusively by the fan loop plasma. The black contours outline the fan loop regions where the initial EM exceeds zero. The accelerated animation presents the simulations for Model 2 in two cases: one including the coronal background (Panels a1--c1) and one including only the fan-loop contribution (Panels a2--c2), over the time interval $t=(1-100)\tau_A$, corresponding to a duration of 710 s.  (An animation of this figure is available in the online article.)    
\label{fig:rdf2}}
\end{figure*}

In this section, we present the simulation results for Model 2, where the propagating fast waves are driven by a periodic force in the $z$-direction, located at ($x_0$, $y_0$, $z_0$) = (0.7, $-0.6$, 0.3). Figure~\ref{fig:rdf2}(a1)-(c1) displays the relative running difference of EM distributions at $t = 100\tau_A$, projected onto the $xy$-, $yz$-, and $xz$-planes. The waves propagate primarily northwestward in the on-disk view (panel a1) and upward into the higher corona in the limb views (panels b1 and c1). Overall, the wave behavior in Model 2 is similar to that in Model 1, with only minor differences. For example, the amplitude contrast between waves propagating along loop-L1 and loop-L3 is reduced in Model 2 (panel a1), while the southward- and upward-propagating waves are more prominent in panel (b1) compared to Model 1.

For the same reason described in Section~\ref{ssc:eem}, we compute the EM distributions contributed solely by the modeled fan loops and present the result at $t = 100\tau_A$ in Figure~\ref{fig:rdf2}(a2)-(d2). The wave patterns in both the on-disk and limb views in Model 2 are overall consistent with those in Model 1. The main distinction is that the amplitude contrast between waves along loop-L1 and loop-L3 is weaker in Model 2 (panel a2), which is consistent with the case that includes the coronal background.

\begin{figure*}
\gridline{\fig{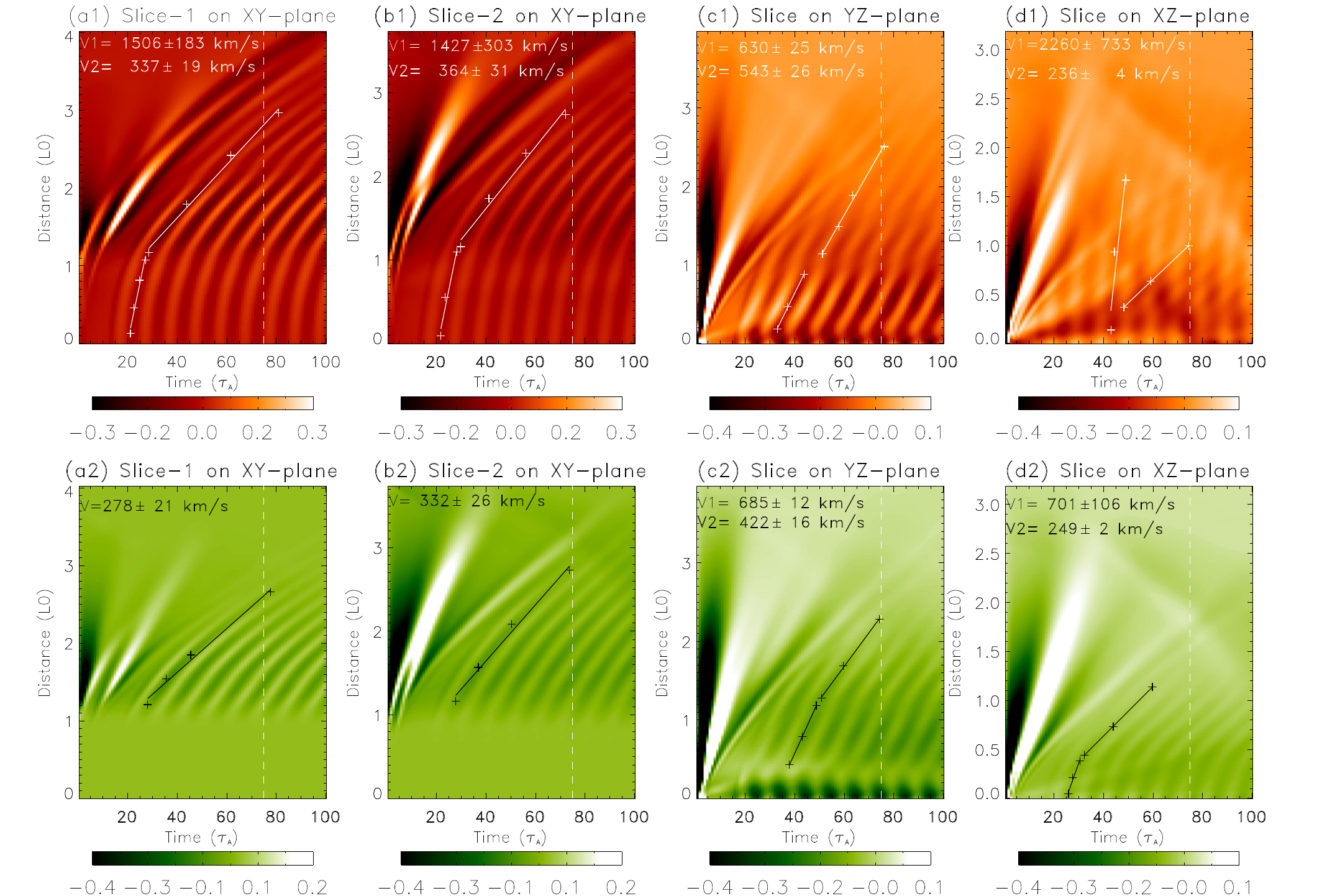}{0.8\textwidth}{}}
\gridline{\fig{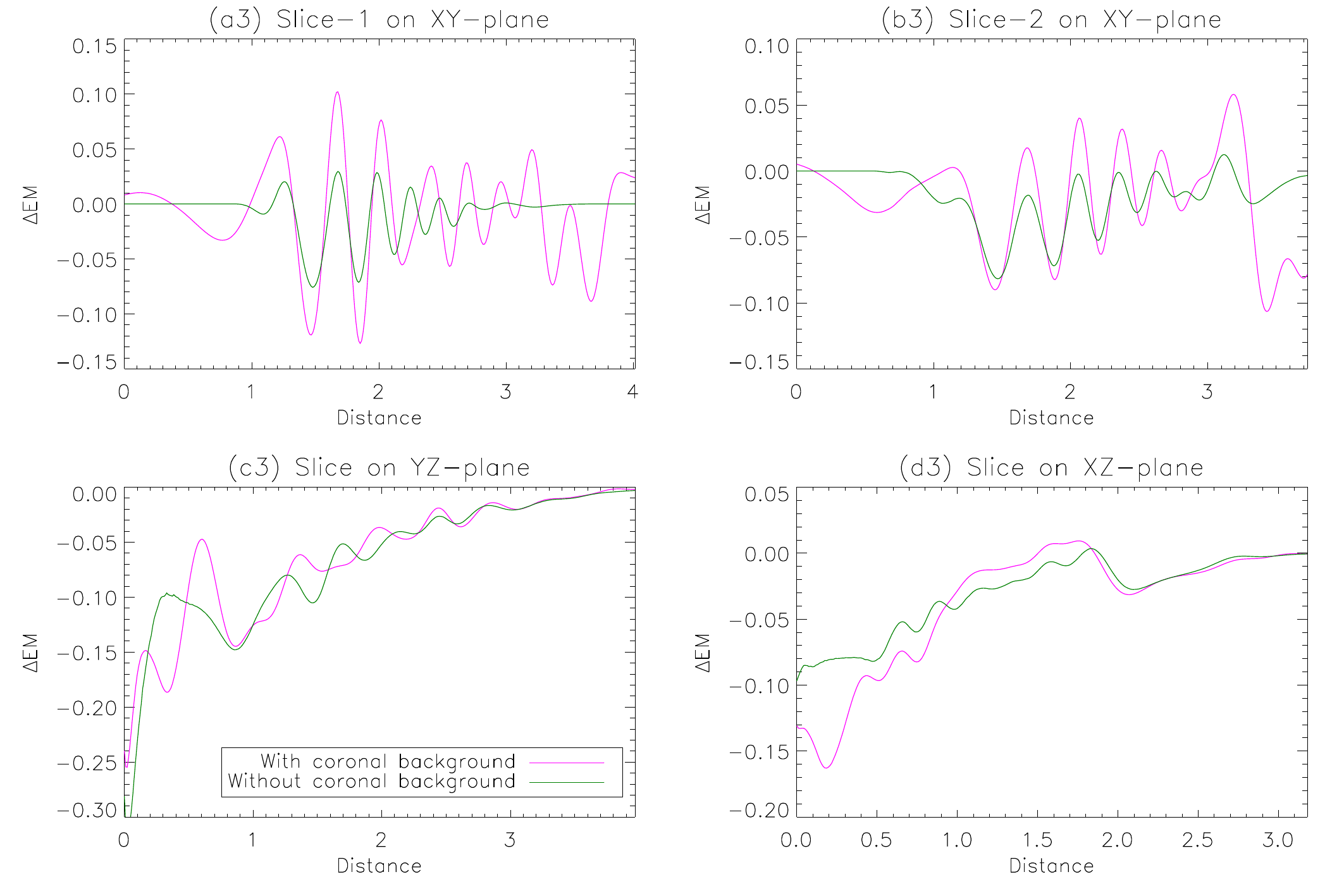}{0.8\textwidth}{}}  
\caption{Panels (a1)-(d1): Time-distance maps of the EM in running difference along the slices shown in Figure~\ref{fig:rdf2}. Slice-1 and Slice-2, used to generate panels (a1) and (b1), respectively, are indicated as S1 and S2 in Figure~\ref{fig:rdf2}(a1). Panels (a2)-(d2): Same as the top panels, but for the EM distributions contributed exclusively by the fan loop plasma. Panels (a3)-(d3): Spatial profiles of the EM in running difference at $t=75\tau_A$ for the vertical cuts shown in the above panels. (a3) Along Slice-1 and (b3) along Slice-2 in the $xy$ plane, (c3) along a slice in the $yz$ plane, and (d3) along a slice in the $xz$ plane. The purple and green lines represent the cases with and without the coronal background included, respectively.     \label{fig:tds2}}
\end{figure*}

The top two rows of Figure~\ref{fig:tds2} present the time-distance maps of EM along four slices in three projected planes. To facilitate comparison between Models~1 and 2, the slices are selected along the same paths as in Model 1. Panels (a1) and (b1) show that the waves propagate from distance $s = 0$ to $\sim 1$ at very high speeds, measured at $1506 \pm 183$ km s$^{-1}$ and $1427 \pm 303$ km s$^{-1}$, respectively. For distances $s \gtrsim 1$, the wave speeds decrease markedly, with estimated values of $337 \pm 19$ km s$^{-1}$ and $364 \pm 31$ km s$^{-1}$, respectively. Panels (a2) and (b2) indicate that the waves propagate at lower speeds, measured as $278 \pm 21$ km s$^{-1}$ and $332 \pm 26$ km s$^{-1}$, respectively. These results for Model 2 demonstrate that the wave behaviors, for both cases with and without the coronal background, when viewed from the top, are broadly consistent with those obtained for Model 1.

In contrast, panels (c1) and (c2) show that, when viewed along the $x$-direction, the wave behaviors differ markedly from those in Model 1 (cf. panels (c1) and (c2) in Figure~\ref{fig:tds1}). In the case with coronal background, the propagation speeds are estimated at $630 \pm 25$ km s$^{-1}$ for $s = 0-1$ and $543 \pm 26$ km s$^{-1}$ for $s > 1$. In the case without coronal background, the speeds are measured as $685 \pm 12$ km s$^{-1}$ for $s = 0.3-1.2$ and $422 \pm 16$ km s$^{-1}$ for $s>1.2$. This difference can be attributed to the weaker amplitude contrast between waves propagating along loop-L1 and loop-L3 in Model 2 compared to Model 1, as discussed above.

Panels (d1) and (d2) show that, when viewed along the $y$-direction, the wave behaviors are similar to those in Model 1, displaying the superposition of a higher-speed component with a slower-speed one. In the case with coronal background, the propagation speeds are estimated at $2260 \pm 733$ km s$^{-1}$ and $236 \pm 4$ km s$^{-1}$. In the case without coronal background, the speeds are $701 \pm 106$ km s$^{-1}$ for $s = 0$–0.5 and $249 \pm 2$ km s$^{-1}$ for $s > 0.5$.

Figure~\ref{fig:tds2}(a3)-(d3) shows the EM profiles along the four slices analyzed above at time $t=75 \tau_A$, for both cases with and without the coronal background. Panels (a3) and (b3) indicate that the wave propagates at a much higher speed in the first wavelength ($s=0$ to $\sim1.2$) near the driving source, roughly four times faster than in the subsequent wavelengths. In the case including the coronal background, the waves display larger amplitudes, propagate faster, and decay more slowly than in the case considering only loop emissions. Panels (c3) and (d3) further demonstrate that, when viewed from the side, the waves exhibit relatively smaller amplitudes and appear to damp more rapidly than when viewed from above. These characteristics are consistent with those found in Model 1.

It is noted that the amplitudes of EM perturbations along Slice-1 and Slice-2 in Model 2 are several times smaller than those in Model 1 (comparing panels (a3) and (b3) between Figure~\ref{fig:tds1} and Figure~\ref{fig:tds2}). This is primarily because the chosen driver amplitude ($f_0$) in Model 2 is not sufficiently large, due to the computational limitations of the current code.

\section{Summary and Discussion}
\label{sec:dac}

\subsection{Summary}
\label{ssec:sum}

We present observational analysis and 3D numerical MHD modeling of the QFP wave trains associated with an M6.5-class flare and CME on 2015 June 22 in AR 12371. The QFPs originate at a distance of about 130 Mm from the flare source, associated with fanlike coronal loops seen in AIA 171 \AA. They travel at speeds of $1500\pm200$ km~s$^{-1}$, with amplitudes of $\sim2-4\%$ in EUV intensity relative to the background corona.  Wavelet analysis reveals that the periodicities ($\sim2-4$ minutes) of the QFPs are consistent with the pulsations in the flare emissions at AIA 1600 and 304~\AA, and Fermi/GBM 26--50~keV, suggestive of a common origin. Such periodicities, simultaneously detected in UV/EUV and hard X-ray wavelengths--particularly in QPPs that do not exhibit obvious smooth damping--have been interpreted as signatures of the same episodic energy release and particle acceleration associated with flare reconnection \citep[see a comprehensive review by][]{zim21}. 

To understand the observed QFP properties, the MHD simulations were performed in a realistic AR magnetic field using the PFSS extrapolation from the HMI vector magnetogram. Motivated by AIA EUV observations, three dense loops were constructed along magnetic field lines in the initial configuration to explore the possible wave-loop interaction and the effect of loop structures on the wave propagation. Using the MHD model NLRAT including the gravitationally stratified solar coronal atmosphere and non-ideal effects, such as thermal conduction and compressive viscosity, we simulated the excitation and propagation of QFP magnetoacoustic waves by a periodic driver in two cases. In the first case (Model 1), we launched periodic velocity pulses \emph{at the QSL footprint near flare ribbons} to model the effect of periodic energy release there. In the second case (Model 2), we exerted periodic forces in a region \emph{above the magnetic neutral line} to model the interaction of reconnection-generated plasmoids in the flare with the ambient magnetic field. 

By analyzing the evolution of physical parameters of the excited waves and quantitatively comparing the wave behaviors in the synthetic emission measures between the two cases -- with and without inclusion of the coronal background to explore its influence on wave visibility -- we summarize the main results as follows:

\begin{enumerate}
\item  In Model 1, the excited fast-mode waves exhibit clear directionality, propagating predominantly northwestward and upward. The wave fronts become distorted when crossing or passing through dense loops; however, the overall influence of coronal loops on the wave propagation remains minor.

\item The excited waves propagate at very high speeds ($\sim1200-1800$ km~s$^{-1}$) near the AR core, traveling over a distance of about 100 Mm within a single period. Beyond this distance, their propagation speed subsequently decreases to roughly one-third of the initial value,  consistent with the decrease in fast-mode speeds due to the rapid decrease of the magnetic field strength with distance away from the AR.

\item The coronal background has a significant influence on the observed wave characteristics. The superposition of waves propagating inside and outside the coronal loops (as they contribute to the observed combined emission in optically thin corona) alters the apparent wave properties -- such as amplitude, propagation speed, and phase -- compared with the case including only the propagation inside the loops.

\item Despite the differences in location and direction of the velocity driver, the wave patterns in Model 2 are overall consistent with those in Model 1 for both cases with and without the coronal background. The only notable difference is that the amplitude contrast among waves propagating along the three loops in Model 2 is weaker than in Model 1.

\item Our simulations indicate that thermal conduction and compressive viscosity have negligible effects on the damping of fast-mode waves under the given coronal conditions, consistent with theoretical predictions \citep{port94}. The weak decrease in wave amplitude with distance observed in the model is primarily a geometric effect, resulting from the spreading of wave energy over an expanding volume as the wavefront becomes increasingly spherical away from the source, and may also involve enhanced resistive dissipation of currents \citep{ofm25}.
\end{enumerate}

\subsection{Discussions}
\label{ssec:disc}

\subsubsection{QFP Energy Flux Estimates}
\label{sssec:energy}

The energy flux carried by QFP wave trains has been estimated in previous studies using the kinetic energy of the perturbed plasma \citep{liu11, shen13, ofm18, zhou21}, yielding values typically in the range of $0.1$–$4.0\times10^5~{\rm erg~cm^{-2}~s^{-1}}$. Since the perturbed velocity, $\delta v$, cannot be directly measured from AIA observations, approximation based on the linear wave theory is adopted, $\delta v \ge \mathbf{k}\!\cdot\!\mathbf{v} = (\delta\rho/\rho)\,v_{\rm ph}$ and $\delta\rho/\rho = \delta I/(2I)$, where $\mathbf{k}\!\cdot\!\mathbf{v}$ denotes the velocity component of the perturbed plasma along the wave propagation direction, and $\delta\rho/\rho$ and $\delta I/I$ are the relative perturbation amplitudes of density and emission intensity, respectively. This leads to an approximate expression for the energy flux based on the WKB approximation:
\begin{equation}
F = \frac{1}{2} \rho (\delta v)^2 v_{\rm ph} \ge \frac{1}{8} \rho \left( \frac{\delta I}{I} \right)^2 v_{\rm ph}^3,
\label{eq:wef}
\end{equation}
which provides a lower-limit estimate. The equality holds when the wave propagates across the magnetic field (i.e., $\mathbf{k}\!\perp\!\mathbf{B}$), corresponding to the condition of maximum density perturbation, $\delta\rho/\rho = \delta B/B$, where the wave signatures are most visible in EUV observations. Therefore, the prominent QFP signals detected in AIA images are expected to originate from locations where $\mathbf{k}\!\perp\!\mathbf{B}$ is approximately satisfied, implying that the estimated lower-limit energy fluxes are likely close to their true values.

For the QFP event analyzed here, by adopting $\delta I / I = 3\%$ and $v_{\rm ph} = 1500~{\rm km~s^{-1}}$ from the measurements, and assuming an electron density for the fan loops of $n_e \approx 1\times10^9~{\rm cm^{-3}}$ \citep{bro11}, we obtain an energy flux of $F\gtrsim6.3\times10^5 ~{\rm erg~cm^{-2}~s^{-1}}$.  This estimated energy flux is comparable to those reported for many previous events using similar methods \citep[e.g.,][]{liu11,ofm18,shen13,shen22,zhou21}.

To facilitate comparison with the simulations, we evaluate the average wave energy flux at location $P_{\rm i2}$ during the interval $t = 55$–$100~\tau_A$ in Model 1 (see Figure~\ref{fig:ncut}C). From the simulation data, the average number density is $n = 5.4\times10^8~{\rm cm^{-3}}$, the velocity perturbation amplitude is $\delta v = 5.4~{\rm km~s^{-1}}$, and the relative density perturbation is $\delta\rho / \rho = 3\%$, corresponding to $\delta I / I \approx 6\%$. Taking the phase speed as $v_{\rm ph} = 350~{\rm km~s^{-1}}$, which represents the mean propagation speed for $s > 1$ (see Figure~\ref{fig:tds1}~a1–c1), and using $F_k=\rho(\delta{v})^2 v_{\rm ph}/2$, we obtain $F_k=4.6\times10^3 ~{\rm erg~cm^{-2}~s^{-1}}$. Moreover, from the amplitude of magnetic field perturbations, $\delta{B}=0.07$ G, and using $F_m=((\delta{B})^2/8\pi) v_{\rm ph}$, we estimate the magnetic energy flux to be $F_m=6.8\times10^3 ~{\rm erg~cm^{-2}~s^{-1}}$.  

We find that the wave energy estimated from both velocity (wave kinetic energy) and magnetic field perturbations in the simulations is about two orders of magnitude lower than that inferred from observations, despite their comparable amplitudes in density perturbations. This discrepancy arises mainly because the modeled phase speed is about 4.3 times smaller than the observed value ($1500/350 \sim 4.3$), which results in a reduction of the wave energy flux by approximately a factor of 80 due to its $v_{\rm ph}^3$ dependence (Equation~\ref{eq:wef}). Assuming $V_f \approx V_A = B/\sqrt{4\pi\rho}$ and $\rho \approx n_e m_p$ with $n_e = 1\times10^9~\mathrm{cm^{-3}}$, we estimate the coronal magnetic field as $B \approx 22~\mathrm{G}$ from the observed QFP phase speed and $B \approx 5~\mathrm{G}$ from the modeled phase speed. These values imply that the PFSS model used here may substantially underestimate the coronal magnetic field strength. Potential-field extrapolations are well known to yield lower field magnitudes than NLFFF and MHD-based approaches, which typically reproduce observed EUV loops more accurately 
\citep[e.g.,][]{rosa09, wieg12, asch13, warr18}. In addition, the PFSS model used in this study employs a synchronic magnetic map \citep{sun18} as its boundary condition, which may further lead to an underestimate of the extrapolated coronal magnetic field strengths, because its spatial resolution is significantly degraded relative to the native HMI pixel size (by a factor of about 16 at the disk center).
 
Specifically, the observed fan loops may be different in magnetic geometry from the simulated ones. The observed fan loops may be lower-lying or more strongly inclined relative to the vertical than the modeled loops, resulting in stronger magnetic fields and thus higher Alfv\'{e}n and fast-mode speeds. Therefore, accurate measurements of the 3D magnetic geometry of coronal structures are crucial for constraining the modeled AR magnetic fields. Such constraints would improve the realism of QFP wave train modeling and lead to more accurate 
 estimates of their energy flux.
                              
Our simulation results can also be used to test the relation $\delta v \ge (\delta \rho / \rho) v_{\rm ph}$ derived from linear wave theory. For the perturbed plasma at location $P_{2i}$, where $\mathbf{k}\!\perp\!\mathbf{B}$ approximately holds, taking $v_{\rm ph}=350$ km~s$^{-1}$ and $\delta \rho / \rho = 3\%$, we estimate $\delta v \gtrsim 11$ km~s$^{-1}$, about twice the value directly measured from the simulations. This indicates that the wave energy flux of the observed QFPs estimated using Equation~(\ref{eq:wef}) may be substantially overestimated, given that the simulations provide a more accurate physical representation. It should be noted that Equation~(\ref{eq:wef}) is derived under the assumption of a uniform magnetic field, which may be valid in the quiet-Sun corona but can introduce significant deviations in the highly structured AR corona. This, in turn, highlights the importance of realistic 3D MHD simulations in advancing coronal seismology for diagnosing unknown coronal parameters.

 Although the estimated energy flux carried by QFP wave trains is  modest compared to typical heating requirements for the AR coronal plasma, their interaction with randomly structured corona may still produce important localized effects. In particular, propagation through regions containing magnetic nulls, QSLs, and other topological gradients can enhance wave-plasma coupling, promote wave energy dissipation (involving mode conversion; \citealp[e.g.,][]{LO14,chen16,pant17}), and potentially modulate or trigger magnetic reconnection \citep[e.g.,][]{mond24,sriv25,liak25}. Such processes are critical to coronal energy transport and may contribute significantly to the heating of the quiet-Sun corona.


\subsubsection{Excitation Mechanisms of QFPs}
\label{sssec:exc}

 In this study, the observed periodicities of the QFPs are consistent with those of the flare emission pulsations, 
 suggesting that they likely originate from magnetic reconnection. Previous MHD simulations have demonstrated that intermittently formed plasmoids in current sheets can generate fast-mode waves through their interaction with the ambient magnetic fields \citep{yang15, taks16, mond24,mond25, liak25}. In Model~2, a periodic driver placed above the magnetic neutral line -- where the flare energy release was observed -- can excite fast-mode waves whose propagation directions are more consistent with the observations, further supporting this scenario.  \citet{mond25} reported that the interaction of reconnection-generated plasmoids with the ends of Y-shaped separatrices can produce slow-mode waves propagating inside the QSLs, in addition to the QFPs propagating outward into the ambient corona. We have not found clear evidence of such slow mode waves in our simulations, most likely due to insufficient spatial resolution to resolve these small-scale structures within the QSLs. However, this aspect lies beyond the scope of the current study, which focuses on large-scale QFP wave trains, and does not affect our conclusions.

On the other hand, our observations and magnetic topology of the analyzed AR do not support the generation of QFP wave trains by a localized, impulsive driver at the footpoints of fan loops through the dispersion-evolution mechanism \citep[e.g.,][]{pas13}. If the impulsive energy release directly heats the associated loops, then loops hosting QFPs would not be expected to appear in the cool AIA 171 \AA\ channel.  However, we cannot exclude the possibility that the observed QFPs result from dispersive evolution and leakage when an impulsive driver is located within a closed flare loop in the AR core \citep[e.g.,][]{nist14,pas17,shi25,shi26} or in a flare current sheet \citep[e.g][]{jel12,mes14}.

\subsubsection{Observing-passband Dependent QFP Bahaviors}
\label{sssec:passband}

The occurrences of QFP wave trains are typically associated with flares. 
While QFPs can be seen close to the corresponding flare kernels under favorable observational conditions \citep[e.g.,][]{liu11,liu12,ofm18}, their initial appearances \citep[e.g.,][]{Liu10,liuw16, shen12,shen18a, miao21} are often located more than 100~Mm away (133~Mm for the event under study). This distance, on the order of 1--2 QFP wavelengths, likely represents the first sign of visibility (or instrumental detection) rather than the actual formation locations of the waves in those events. 
In fact, our simulations demonstrate that the wave pulses within such a distance are difficult to discern, because of small wave amplitudes owing to wave energy conservation in the presence of strong magnetic fields and high fast-mode speeds in the AR core.%
\footnote{A similar effect is present in global EUV (EIT) waves that can ``disappear" while traversing an AR because of significantly reduced wave amplitudes there, again as a result of wave energy conservation, and then reappearing from the other side of the AR \citep[e.g.,][]{OT02, LiT12}.
Also note that coincidentally, global EUV waves tend to first appear at a similar distance on the order of 100~Mm from the eruption center for a different reason than that noted here for QFPs
\citep[e.g., see][their Section 3.10]{LO14}.}
QFPs can be increasingly visible when they propagate slightly away from the AR, where the decreasing magnetic field strength and fast-mode speed lead to their amplitudes to grow and become more readily detectable, despite the geometric expansion described above. This distance also avoids the observational confusion from strong EUV emission near the flare kernel, thus allowing QFP wave pulses to stand out.


Moreover, the simulations suggest that the commonly observed association of QFP wave trains with coronal fan loops \citep[e.g.,][]{liu11, miao19, miao21}, best seen in AIA 171 \AA, may be attributed to the thermal response characteristics of AIA passbands, which affects the detectability of QFP waves. This 171 \AA\ channel is most sensitive to relatively cool ($T_e < 1$ MK) coronal structures--notably those fan loops rooted at the AR edge--but not to hot plasma such as the AR core ($T_e \approx 2$--4~MK) or the warmer QS ($T_e \approx 1$--2~MK). This makes the fan loops stand out prominently in AIA 171~\AA\ images against the surrounding AR core and QS in the coronal background (including contributions from foreground and background coronal plasma), where there is very little 171~\AA\ emission. 
Consequently, QFPs are best seen and appear to be spatially confined within those cool fan loops, but remain barely visible elsewhere. 




In this scenario, QFP wave trains should in principle be detectable in other AIA channels such as 193 and 211 \AA, because the propagation and dissipation of fast-mode waves are theoretically nearly independent of plasma temperature in low-$\beta$ conditions. Since the 193 and 211 \AA\ channels are more sensitive to the warmer ($T_e = 1-2$ MK) corona in QS regions and the fast-mode speed distribution is more uniform there, the QFPs would be expected to appear more prominently there in these two channels. This conjecture is supported by our simulations, which show that the case including the coronal background exhibits larger wave amplitudes and weaker damping compared to the case considering only loop emissions.

However, analysis of the AIA 193 \AA\ observations revealed no detectable signatures of QFPs in the event under study. In fact, previous observations have shown that QFPs are most frequently detected in the AIA 171 \AA\ channel, while detections in the 193 or 211 \AA\ channels are rare \citep[e.g.,][]{Liu10,liuw16, shen13}. Fast-mode wave trains seen in the latter usually appear further away from the AR core, and can follow the global EIT wave front to travel to much greater distances and at lower speeds than their 171~\AA\ counterpart. Thus they have been called ``EIT wave trains" and are conjectured to be the continuation of QFPs into the QS region \citep[][their Section 4.3]{LO14}, which is now supported by our simulations.


These results suggest that the preferential detection of QFPs in AIA 171 \AA\ is likely due to its photon response efficiency being significantly higher (by at least an order of magnitude) than that of other AIA channels \citep{boer12}, making it particularly sensitive to small intensity variations, typically at the 1$-$5\% level for QFPs \citep{liuw16}. In addition, the cool 171~\AA\ fan loops rooted in ARs usually appear well in isolation with very little coronal background, as noted above, which allows QFPs in these loops to readily stand out with decent contrast and S/N. This is not the case for 193 and 211~\AA\ channels, whose QFP signals in these structures are smeared by LOS integration of foreground and background optically thin emission from the QS surrounding the AR.

\subsubsection{Other Effects on QFP Detectability and Visibility}
\label{sssec:other}

Our simulations also indicate that  
the visibility of QFPs depends on the angle between the LOS and the wave propagation direction. The waves are most readily detected when the LOS is perpendicular to the direction of propagation, allowing longer LOS-integration-path lengths parallel to the wave arc-shaped front and thus increased contrast. However, when multiple coronal loops are present along the LOS, the superposition of QFPs propagating in different structures can lead to partial cancellation due to differences in their propagation speeds and phases and the LOS integration in the optically thin corona, resulting in significantly reduced wave amplitudes that are difficult to detect. This effect agrees with the initial finding of \citet{Downs15} based on the simulation case reported in \citet{Downs21}.


MHD simulations have suggested that the properties of QFP wave trains$-$such as their propagation speed, direction, and amplitude$-$depend on the location, directionality, and magnitude of the flare, as well as on the AR magnetic geometry \citep{ofm11, ofm18, ofm25}. In our simulations, the QFPs excited by two periodic drivers, differing in both position and driving direction as motivated by AIA observations, exhibit overall similar wave patterns. The reason could be that, owing to the very high fast-mode speeds in the AR, drivers located at different positions within the core are effectively close to each other on timescales of the characteristic coronal Alfv\'{e}n speed (typically 1000$-$3000 km~s$^{-1}$; \citealt{nak01,asch04,LO14}). It is noteworthy, however, that the dominant propagation direction of the modeled waves does not agree with the observed one. This discrepancy may arise from the use of a potential field model as the initial magnetic configuration of the AR, which differs significantly from the real field where strong magnetic shear typically exists near the flare source (see Figure~\ref{fig:bimg}).


\label{sssec:damp}



While physical damping has been present in previous modeling results \citep[e.g.,][]{nist14,ofm25}, our simulations demonstrate that the modeled QFP waves exhibit much weaker damping than seen in observations during propagation. This appears inconsistent with observations, in which QFP wave trains often decay within a few wavelengths or rapidly vanish beyond a certain distance \citep[e.g.,][]{liu11,liu12}. As discussed above, such observed strong damping may not necessarily imply true physical dissipation. The observed decay or disappearance could, at least in part, be attributed to the combination of various observational effects, such as the relative angle between the LOS and the QFP propagation direction that can change rapidly, the multi-thermal and structured (including gravitational stratification) nature of the corona plasma, and the temperature-dependent instrument sensitivity of the observing passbands. 

The exact extents and relative contributions of such observational effects and physical damping in QFPs remain an open question. Future instruments, such as the Multi-slit Solar Explorer \citep[MUSE;][]{DePont20} mission and the EUV High-throughput Spectroscopic Telescope (EUVST) onboard the Solar-C mission \citep{shim20}, with improved sensitivity, spatiotemporal resolution, as well as spectroscopic measurements (of Doppler velocities), can shed critical light on this and other open questions of QFPs.
\vspace{5mm}
\noindent 
The authors acknowledge support by NASA grants 80NSSC21K1687 and 80NSSC22K0527. The work of T.W. and L.O. was also supported by NASA grant 80NSSC22K0755 and NASA's GSFC through Cooperative Agreement 80NSSC21M0180 to Catholic University of America, Partnership for Heliophysics and Space Environment Research (PHaSER). 
M.J. and W.L. acknowledges support from NASA’s SDO/AIA contract (NNG04EA00C) to LMSAL.  We also thank the anonymous referee for their constructive comments which have improved the clarity and contents of the paper. Computational resources supporting this work were provided by the NASA High-End Computing (HEC) Program through the NASA Advanced Supercomputing (NAS) Division at Ames Research Center.


\bibliography{wang_apj}{}

\begin{thebibliography}{}
\expandafter\ifx\csname natexlab\endcsname\relax\def\natexlab#1{#1}\fi
\providecommand{\url}[1]{\href{#1}{#1}}
\providecommand{\dodoi}[1]{doi:~\href{http://doi.org/#1}{\nolinkurl{#1}}}
\providecommand{\doeprint}[1]{\href{http://ascl.net/#1}{\nolinkurl{http://ascl.net/#1}}}
\providecommand{\doarXiv}[1]{\href{https://arxiv.org/abs/#1}{\nolinkurl{https://arxiv.org/abs/#1}}}

\bibitem[{{Aschwanden}(2004)}]{asch04}
{Aschwanden}, M.~J. 2004, {Physics of the Solar Corona. An Introduction}
  (Berlin, Heidelberg: Springer)

\bibitem[{{Aschwanden}(2013)}]{asch13}
---. 2013, \apj, 763, 115, \dodoi{10.1088/0004-637X/763/2/115}

\bibitem[{Ashfield {et~al.}(2026)Ashfield, Polito, L{\"o}rin{\v{c}}{\'\i}k,
  De~Pontieu, Chintzoglou, Bose, Freij, Rouppe van~der Voort, Joshi, \&
  Faber}]{Ashf26}
Ashfield, W.~I., Polito, V., L{\"o}rin{\v{c}}{\'\i}k, J., {et~al.} 2026, Nature
  Astronomy, 10, 54, \dodoi{10.1038/s41550-025-02706-4}

\bibitem[{{Boerner} {et~al.}(2012){Boerner}, {Edwards}, {Lemen}, {Rausch},
  {Schrijver}, {Shine}, {Shing}, {Stern}, {Tarbell}, {Title}, {Wolfson},
  {Soufli}, {Spiller}, {Gullikson}, {McKenzie}, {Windt}, {Golub}, {Podgorski},
  {Testa}, \& {Weber}}]{boer12}
{Boerner}, P., {Edwards}, C., {Lemen}, J., {et~al.} 2012, \solphys, 275, 41,
  \dodoi{10.1007/s11207-011-9804-8}

\bibitem[{{Brooks} {et~al.}(2011){Brooks}, {Warren}, \& {Young}}]{bro11}
{Brooks}, D.~H., {Warren}, H.~P., \& {Young}, P.~R. 2011, \apj, 730, 85,
  \dodoi{10.1088/0004-637X/730/2/85}

\bibitem[{{Chen} {et~al.}(2016){Chen}, {Fang}, {Chandra}, \&
  {Srivastava}}]{chen16}
{Chen}, P.~F., {Fang}, C., {Chandra}, R., \& {Srivastava}, A.~K. 2016,
  \solphys, 291, 3195, \dodoi{10.1007/s11207-016-0920-3}

\bibitem[{{De Pontieu} {et~al.}(2020){De Pontieu}, {Mart{\'\i}nez-Sykora},
  {Testa}, {Winebarger}, {Daw}, {Hansteen}, {Cheung}, \& {Antolin}}]{DePont20}
{De Pontieu}, B., {Mart{\'\i}nez-Sykora}, J., {Testa}, P., {et~al.} 2020, \apj,
  888, 3, \dodoi{10.3847/1538-4357/ab5b03}

\bibitem[{{De Rosa} {et~al.}(2009){De Rosa}, {Schrijver}, {Barnes}, {Leka},
  {Lites}, {Aschwanden}, {Amari}, {Canou}, {McTiernan}, {R{\'e}gnier},
  {Thalmann}, {Valori}, {Wheatland}, {Wiegelmann}, {Cheung}, {Conlon},
  {Fuhrmann}, {Inhester}, \& {Tadesse}}]{rosa09}
{De Rosa}, M.~L., {Schrijver}, C.~J., {Barnes}, G., {et~al.} 2009, \apj, 696,
  1780, \dodoi{10.1088/0004-637X/696/2/1780}

\bibitem[{{Del Zanna} \& {Mason}(2003)}]{del03}
{Del Zanna}, G., \& {Mason}, H.~E. 2003, \aap, 406, 1089,
  \dodoi{10.1051/0004-6361:20030791}

\bibitem[{{Demoulin} {et~al.}(1997){Demoulin}, {Bagala}, {Mandrini}, {Henoux},
  \& {Rovira}}]{dem97}
{Demoulin}, P., {Bagala}, L.~G., {Mandrini}, C.~H., {Henoux}, J.~C., \&
  {Rovira}, M.~G. 1997, \aap, 325, 305

\bibitem[{{D{\'e}moulin} {et~al.}(1996){D{\'e}moulin}, {Priest}, \&
  {Lonie}}]{dem96}
{D{\'e}moulin}, P., {Priest}, E.~R., \& {Lonie}, D.~P. 1996, \jgr, 101, 7631,
  \dodoi{10.1029/95JA03558}

\bibitem[{{Dennis} \& {Zarro}(1993)}]{den93}
{Dennis}, B.~R., \& {Zarro}, D.~M. 1993, \solphys, 146, 177,
  \dodoi{10.1007/BF00662178}

\bibitem[{{Downs} {et~al.}(2015){Downs}, {Liu}, {Torok}, {Linker}, {Mikic}, \&
  {Ofman}}]{Downs15}
{Downs}, C., {Liu}, W., {Torok}, T., {et~al.} 2015, in AGU Fall Meeting
  Abstracts, Vol. 2015, AGU Fall Meeting Abstracts, SH22A--06

\bibitem[{{Downs} {et~al.}(2021){Downs}, {Warmuth}, {Long}, {Bloomfield},
  {Kwon}, {Veronig}, {Vourlidas}, \& {Vr{\v{s}}nak}}]{Downs21}
{Downs}, C., {Warmuth}, A., {Long}, D.~M., {et~al.} 2021, \apj, 911, 118,
  \dodoi{10.3847/1538-4357/abea78}

\bibitem[{{Hu} {et~al.}(2024){Hu}, {Ye}, {Chen}, {Mei}, {Tang}, \&
  {Lin}}]{hu24}
{Hu}, J., {Ye}, J., {Chen}, Y., {et~al.} 2024, \apj, 962, 42,
  \dodoi{10.3847/1538-4357/ad1993}

\bibitem[{{Jel{\'\i}nek} {et~al.}(2012){Jel{\'\i}nek}, {Karlick{\'y}}, \&
  {Murawski}}]{jel12}
{Jel{\'\i}nek}, P., {Karlick{\'y}}, M., \& {Murawski}, K. 2012, \aap, 546, A49,
  \dodoi{10.1051/0004-6361/201219891}

\bibitem[{{Jin} {et~al.}(2022){Jin}, {Cheung}, {DeRosa}, {Nitta}, \&
  {Schrijver}}]{jin22}
{Jin}, M., {Cheung}, M. C.~M., {DeRosa}, M.~L., {Nitta}, N.~V., \& {Schrijver},
  C.~J. 2022, \apj, 928, 154, \dodoi{10.3847/1538-4357/ac589b}

\bibitem[{{Jing} {et~al.}(2017){Jing}, {Liu}, {Cheung}, {Lee}, {Xu}, {Liu},
  {Zhu}, \& {Wang}}]{Jing17}
{Jing}, J., {Liu}, R., {Cheung}, M. C.~M., {et~al.} 2017, \apjl, 842, L18,
  \dodoi{10.3847/2041-8213/aa774d}

\bibitem[{{Jing} {et~al.}(2016){Jing}, {Xu}, {Cao}, {Liu}, {Gary}, \&
  {Wang}}]{Jing16}
{Jing}, J., {Xu}, Y., {Cao}, W., {et~al.} 2016, Scientific Reports, 6, 24319,
  \dodoi{10.1038/srep24319}

\bibitem[{{Kolotkov} {et~al.}(2021){Kolotkov}, {Nakariakov}, {Moss}, \&
  {Shellard}}]{kolot21}
{Kolotkov}, D.~Y., {Nakariakov}, V.~M., {Moss}, G., \& {Shellard}, P. 2021,
  \mnras, 505, 3505, \dodoi{10.1093/mnras/stab1587}

\bibitem[{{Krishna Prasad} {et~al.}(2018){Krishna Prasad}, {Raes}, {Van
  Doorsselaere}, {Magyar}, \& {Jess}}]{krish18}
{Krishna Prasad}, S., {Raes}, J.~O., {Van Doorsselaere}, T., {Magyar}, N., \&
  {Jess}, D.~B. 2018, \apj, 868, 149, \dodoi{10.3847/1538-4357/aae9f5}

\bibitem[{{Li} {et~al.}(2012){Li}, {Zhang}, {Yang}, \& {Liu}}]{LiT12}
{Li}, T., {Zhang}, J., {Yang}, S.-H., \& {Liu}, W. 2012, \apj, 746, 13,
  \dodoi{10.1088/0004-637X/746/1/13}

\bibitem[{{Liakh} \& {Keppens}(2025)}]{liak25}
{Liakh}, V., \& {Keppens}, R. 2025, \aap, 696, A158,
  \dodoi{10.1051/0004-6361/202453300}

\bibitem[{{Linker} {et~al.}(1999){Linker}, {Miki{\'c}}, {Biesecker}, {Forsyth},
  {Gibson}, {Lazarus}, {Lecinski}, {Riley}, {Szabo}, \& {Thompson}}]{link99}
{Linker}, J.~A., {Miki{\'c}}, Z., {Biesecker}, D.~A., {et~al.} 1999, \jgr, 104,
  9809, \dodoi{10.1029/1998JA900159}

\bibitem[{{Liu} {et~al.}(2016{\natexlab{a}}){Liu}, {Kliem}, {Titov}, {Chen},
  {Wang}, {Wang}, {Liu}, {Xu}, \& {Wiegelmann}}]{liur16}
{Liu}, R., {Kliem}, B., {Titov}, V.~S., {et~al.} 2016{\natexlab{a}}, \apj, 818,
  148, \dodoi{10.3847/0004-637X/818/2/148}

\bibitem[{{Liu} {et~al.}(2010){Liu}, {Nitta}, {Schrijver}, {Title}, \&
  {Tarbell}}]{Liu10}
{Liu}, W., {Nitta}, N.~V., {Schrijver}, C.~J., {Title}, A.~M., \& {Tarbell},
  T.~D. 2010, \apjl, 723, L53, \dodoi{10.1088/2041-8205/723/1/L53}

\bibitem[{{Liu} \& {Ofman}(2014)}]{LO14}
{Liu}, W., \& {Ofman}, L. 2014, \solphys, 289, 3233,
  \dodoi{10.1007/s11207-014-0528-4}

\bibitem[{{Liu} {et~al.}(2016{\natexlab{b}}){Liu}, {Ofman}, {Broder},
  {Karlick{\'y}}, \& {Downs}}]{liuw16}
{Liu}, W., {Ofman}, L., {Broder}, B., {Karlick{\'y}}, M., \& {Downs}, C.
  2016{\natexlab{b}}, in American Institute of Physics Conference Series, Vol.
  1720, Solar Wind 14 (AIP), 040010, \dodoi{10.1063/1.4943821}

\bibitem[{{Liu} {et~al.}(2012){Liu}, {Ofman}, {Nitta}, {Aschwanden},
  {Schrijver}, {Title}, \& {Tarbell}}]{liu12}
{Liu}, W., {Ofman}, L., {Nitta}, N.~V., {et~al.} 2012, \apj, 753, 52,
  \dodoi{10.1088/0004-637X/753/1/52}

\bibitem[{{Liu} {et~al.}(2011){Liu}, {Title}, {Zhao}, {Ofman}, {Schrijver},
  {Aschwanden}, {De Pontieu}, \& {Tarbell}}]{liu11}
{Liu}, W., {Title}, A.~M., {Zhao}, J., {et~al.} 2011, \apjl, 736, L13,
  \dodoi{10.1088/2041-8205/736/1/L13}

\bibitem[{{Mann} \& {Veronig}(2023)}]{mann23}
{Mann}, G., \& {Veronig}, A.~M. 2023, \aap, 676, A144,
  \dodoi{10.1051/0004-6361/202245688}

\bibitem[{{McLaughlin} \& {Hood}(2004)}]{mcl04}
{McLaughlin}, J.~A., \& {Hood}, A.~W. 2004, \aap, 420, 1129,
  \dodoi{10.1051/0004-6361:20035900}

\bibitem[{{McLaughlin} \& {Ofman}(2008)}]{mcl08}
{McLaughlin}, J.~A., \& {Ofman}, L. 2008, \apj, 682, 1338,
  \dodoi{10.1086/588799}

\bibitem[{{Mei} {et~al.}(2020){Mei}, {Keppens}, {Cai}, {Ye}, {Xie}, \&
  {Li}}]{mei20}
{Mei}, Z.~X., {Keppens}, R., {Cai}, Q.~W., {et~al.} 2020, \mnras, 493, 4816,
  \dodoi{10.1093/mnras/staa555}

\bibitem[{{M{\'e}sz{\'a}rosov{\'a}} {et~al.}(2014){M{\'e}sz{\'a}rosov{\'a}},
  {Karlick{\'y}}, {Jel{\'\i}nek}, \& {Ryb{\'a}k}}]{mes14}
{M{\'e}sz{\'a}rosov{\'a}}, H., {Karlick{\'y}}, M., {Jel{\'\i}nek}, P., \&
  {Ryb{\'a}k}, J. 2014, \apj, 788, 44, \dodoi{10.1088/0004-637X/788/1/44}

\bibitem[{{Miao} {et~al.}(2021){Miao}, {Li}, {Yuan}, {Jiang}, {Elmhamdi},
  {Zhao}, \& {Anfinogentov}}]{miao21}
{Miao}, Y., {Li}, D., {Yuan}, D., {et~al.} 2021, \apjl, 908, L37,
  \dodoi{10.3847/2041-8213/abdfce}

\bibitem[{{Miao} {et~al.}(2019){Miao}, {Liu}, {Shen}, {Li}, {Abidin},
  {Elmhamdi}, \& {Kordi}}]{miao19}
{Miao}, Y.~H., {Liu}, Y., {Shen}, Y.~D., {et~al.} 2019, \apjl, 871, L2,
  \dodoi{10.3847/2041-8213/aafaf9}

\bibitem[{{Mondal} {et~al.}(2025){Mondal}, {Srivastava}, {Pontin}, \&
  {Priest}}]{mond25}
{Mondal}, S., {Srivastava}, A.~K., {Pontin}, D.~I., \& {Priest}, E.~R. 2025,
  \apj, 989, 222, \dodoi{10.3847/1538-4357/adf18e}

\bibitem[{{Mondal} {et~al.}(2024){Mondal}, {Srivastava}, {Pontin}, {Priest},
  {Kwon}, \& {Yuan}}]{mond24}
{Mondal}, S., {Srivastava}, A.~K., {Pontin}, D.~I., {et~al.} 2024, \apj, 977,
  235, \dodoi{10.3847/1538-4357/ad9022}

\bibitem[{{Nakariakov} \& {Ofman}(2001)}]{nak01}
{Nakariakov}, V.~M., \& {Ofman}, L. 2001, \aap, 372, L53,
  \dodoi{10.1051/0004-6361:20010607}

\bibitem[{{Nakariakov} {et~al.}(2024){Nakariakov}, {Zhong}, {Kolotkov},
  {Meadowcroft}, {Zhong}, \& {Yuan}}]{nak24}
{Nakariakov}, V.~M., {Zhong}, S., {Kolotkov}, D.~Y., {et~al.} 2024, Reviews of
  Modern Plasma Physics, 8, 19, \dodoi{10.1007/s41614-024-00160-9}

\bibitem[{{Nistic{\`o}} {et~al.}(2014){Nistic{\`o}}, {Pascoe}, \&
  {Nakariakov}}]{nist14}
{Nistic{\`o}}, G., {Pascoe}, D.~J., \& {Nakariakov}, V.~M. 2014, \aap, 569,
  A12, \dodoi{10.1051/0004-6361/201423763}

\bibitem[{{Ofman} \& {Liu}(2018)}]{ofm18}
{Ofman}, L., \& {Liu}, W. 2018, \apj, 860, 54, \dodoi{10.3847/1538-4357/aac2e8}

\bibitem[{{Ofman} {et~al.}(2011){Ofman}, {Liu}, {Title}, \&
  {Aschwanden}}]{ofm11}
{Ofman}, L., {Liu}, W., {Title}, A., \& {Aschwanden}, M. 2011, \apjl, 740, L33,
  \dodoi{10.1088/2041-8205/740/2/L33}

\bibitem[{{Ofman} \& {Thompson}(2002)}]{OT02}
{Ofman}, L., \& {Thompson}, B.~J. 2002, \apj, 574, 440, \dodoi{10.1086/340924}

\bibitem[{{Ofman} \& {Wang}(2022)}]{ofm22}
{Ofman}, L., \& {Wang}, T. 2022, \apj, 926, 64,
  \dodoi{10.3847/1538-4357/ac4090}

\bibitem[{{Ofman} {et~al.}(2025){Ofman}, {Wang}, {Sun}, \& {Jin}}]{ofm25}
{Ofman}, L., {Wang}, T., {Sun}, X., \& {Jin}, M. 2025, \apj, 994, 25,
  \dodoi{10.3847/1538-4357/ae1036}

\bibitem[{{Pant} {et~al.}(2017){Pant}, {Tiwari}, {Yuan}, \&
  {Banerjee}}]{pant17}
{Pant}, V., {Tiwari}, A., {Yuan}, D., \& {Banerjee}, D. 2017, \apjl, 847, L5,
  \dodoi{10.3847/2041-8213/aa880f}

\bibitem[{{Pascoe} {et~al.}(2017){Pascoe}, {Goddard}, \& {Nakariakov}}]{pas17}
{Pascoe}, D.~J., {Goddard}, C.~R., \& {Nakariakov}, V.~M. 2017, \apjl, 847,
  L21, \dodoi{10.3847/2041-8213/aa8db8}

\bibitem[{{Pascoe} {et~al.}(2013){Pascoe}, {Nakariakov}, \&
  {Kupriyanova}}]{pas13}
{Pascoe}, D.~J., {Nakariakov}, V.~M., \& {Kupriyanova}, E.~G. 2013, \aap, 560,
  A97, \dodoi{10.1051/0004-6361/201322678}

\bibitem[{{Porter} {et~al.}(1994){Porter}, {Klimchuk}, \& {Sturrock}}]{port94}
{Porter}, L.~J., {Klimchuk}, J.~A., \& {Sturrock}, P.~A. 1994, \apj, 435, 502,
  \dodoi{10.1086/174831}

\bibitem[{{Provornikova} {et~al.}(2018){Provornikova}, {Ofman}, \&
  {Wang}}]{prov18}
{Provornikova}, E., {Ofman}, L., \& {Wang}, T. 2018, Adv. Space Res., 61, 645,
  \dodoi{10.1016/j.asr.2017.07.042}

\bibitem[{{Qiu} {et~al.}(2012){Qiu}, {Liu}, \& {Longcope}}]{qiu12}
{Qiu}, J., {Liu}, W.-J., \& {Longcope}, D.~W. 2012, \apj, 752, 124,
  \dodoi{10.1088/0004-637X/752/2/124}

\bibitem[{{Reale}(2026)}]{real26}
{Reale}, F. 2026, arXiv e-prints, arXiv:2602.03499,
  \dodoi{10.48550/arXiv.2602.03499}

\bibitem[{{Selwa} {et~al.}(2011{\natexlab{a}}){Selwa}, {Ofman}, \&
  {Solanki}}]{selw11a}
{Selwa}, M., {Ofman}, L., \& {Solanki}, S.~K. 2011{\natexlab{a}}, \apj, 726,
  42, \dodoi{10.1088/0004-637X/726/1/42}

\bibitem[{{Selwa} {et~al.}(2011{\natexlab{b}}){Selwa}, {Solanki}, \&
  {Ofman}}]{selw11b}
{Selwa}, M., {Solanki}, S.~K., \& {Ofman}, L. 2011{\natexlab{b}}, \apj, 728,
  87, \dodoi{10.1088/0004-637X/728/2/87}

\bibitem[{{Shen} \& {Liu}(2012)}]{shen12}
{Shen}, Y., \& {Liu}, Y. 2012, \apj, 753, 53,
  \dodoi{10.1088/0004-637X/753/1/53}

\bibitem[{{Shen} {et~al.}(2018{\natexlab{a}}){Shen}, {Liu}, {Song}, \&
  {Tian}}]{shen18a}
{Shen}, Y., {Liu}, Y., {Song}, T., \& {Tian}, Z. 2018{\natexlab{a}}, \apj, 853,
  1, \dodoi{10.3847/1538-4357/aaa3ff}

\bibitem[{{Shen} {et~al.}(2018{\natexlab{b}}){Shen}, {Song}, \&
  {Liu}}]{shen18b}
{Shen}, Y., {Song}, T., \& {Liu}, Y. 2018{\natexlab{b}}, \mnras, 477, L6,
  \dodoi{10.1093/mnrasl/sly044}

\bibitem[{{Shen} {et~al.}(2018{\natexlab{c}}){Shen}, {Tang}, {Li}, \&
  {Liu}}]{shen18c}
{Shen}, Y., {Tang}, Z., {Li}, H., \& {Liu}, Y. 2018{\natexlab{c}}, \mnras, 480,
  L63, \dodoi{10.1093/mnrasl/sly127}

\bibitem[{{Shen} {et~al.}(2022){Shen}, {Zhou}, {Duan}, {Tang}, {Zhou}, \&
  {Tan}}]{shen22}
{Shen}, Y., {Zhou}, X., {Duan}, Y., {et~al.} 2022, \solphys, 297, 20,
  \dodoi{10.1007/s11207-022-01953-2}

\bibitem[{{Shen} {et~al.}(2013){Shen}, {Liu}, {Su}, {Li}, {Zhang}, {Tian},
  {Zhao}, \& {Elmhamdi}}]{shen13}
{Shen}, Y.~D., {Liu}, Y., {Su}, J.~T., {et~al.} 2013, \solphys, 288, 585,
  \dodoi{10.1007/s11207-013-0395-4}

\bibitem[{{Shi} {et~al.}(2025){Shi}, {Nakariakov}, {Li}, \& {Guo}}]{shi25}
{Shi}, M., {Nakariakov}, V.~M., {Li}, B., \& {Guo}, M. 2025, \apj, 990, 1,
  \dodoi{10.3847/1538-4357/adf647}

\bibitem[{{Shi} {et~al.}(2026){Shi}, {Nakariakov}, {Li}, \& {Guo}}]{shi26}
---. 2026, \apj, 996, 72, \dodoi{10.3847/1538-4357/ae2746}

\bibitem[{{Shimizu} {et~al.}(2020){Shimizu}, {Imada}, {Kawate}, {Suematsu},
  {Hara}, {Tsuzuki}, {Katsukawa}, {Kubo}, {Ishikawa}, {Watanabe}, {Toriumi},
  {Ichimoto}, {Nagata}, {Hasegawa}, {Yokoyama}, {Watanabe}, {Tsuno},
  {Korendyke}, {Warren}, {De Pontieu}, {Boerner}, {Solanki}, {Teriaca},
  {Schuehle}, {Matthews}, {Long}, {Thomas}, {Hancock}, {Reid}, {Fludra},
  {Auch{\`e}re}, {Andretta}, {Naletto}, {Poletto}, \& {Harra}}]{shim20}
{Shimizu}, T., {Imada}, S., {Kawate}, T., {et~al.} 2020, in Society of
  Photo-Optical Instrumentation Engineers (SPIE) Conference Series, Vol. 11444,
  Space Telescopes and Instrumentation 2020: Ultraviolet to Gamma Ray, ed.
  J.-W.~A. {den Herder}, S.~{Nikzad}, \& K.~{Nakazawa}, 114440N,
  \dodoi{10.1117/12.2560887}

\bibitem[{{Spitzer} \& {H{\"a}rm}(1953)}]{spit53}
{Spitzer}, L., \& {H{\"a}rm}, R. 1953, Phys. Rev., 89, 977,
  \dodoi{10.1103/PhysRev.89.977}

\bibitem[{{Srivastava} {et~al.}(2025){Srivastava}, {Mondal}, {Priest},
  {Mishra}, {Pontin}, {Kwon}, {Yuan}, {Murawski}, \& {Asai}}]{sriv25}
{Srivastava}, A.~K., {Mondal}, S., {Priest}, E.~R., {et~al.} 2025, \apj, 984,
  36, \dodoi{10.3847/1538-4357/adc379}

\bibitem[{{Sun}(2018)}]{sun18}
{Sun}, X. 2018, arXiv e-prints, arXiv:1801.04265,
  \dodoi{10.48550/arXiv.1801.04265}

\bibitem[{{Takasao} \& {Shibata}(2016{\natexlab{a}})}]{tak16}
{Takasao}, S., \& {Shibata}, K. 2016{\natexlab{a}}, \apj, 823, 150,
  \dodoi{10.3847/0004-637X/823/2/150}

\bibitem[{{Takasao} \& {Shibata}(2016{\natexlab{b}})}]{taks16}
---. 2016{\natexlab{b}}, \apj, 823, 150, \dodoi{10.3847/0004-637X/823/2/150}

\bibitem[{{Titov}(2007)}]{tit07}
{Titov}, V.~S. 2007, \apj, 660, 863, \dodoi{10.1086/512671}

\bibitem[{{Titov} {et~al.}(2002){Titov}, {Hornig}, \& {D{\'e}moulin}}]{tit02}
{Titov}, V.~S., {Hornig}, G., \& {D{\'e}moulin}, P. 2002, Journal of
  Geophysical Research (Space Physics), 107, 1164, \dodoi{10.1029/2001JA000278}

\bibitem[{{Torrence} \& {Compo}(1998)}]{torr98}
{Torrence}, C., \& {Compo}, G.~P. 1998, Bulletin of the American Meteorological
  Society, 79, 61, \dodoi{10.1175/1520-0477(1998)079<0061:APGTWA>2.0.CO;2}

\bibitem[{{Van Doorsselaere} {et~al.}(2016){Van Doorsselaere}, {Kupriyanova},
  \& {Yuan}}]{door16}
{Van Doorsselaere}, T., {Kupriyanova}, E.~G., \& {Yuan}, D. 2016, \solphys,
  291, 3143, \dodoi{10.1007/s11207-016-0977-z}

\bibitem[{{Vanninathan} {et~al.}(2015){Vanninathan}, {Veronig}, {Dissauer},
  {Madjarska}, {Hannah}, \& {Kontar}}]{van15}
{Vanninathan}, K., {Veronig}, A.~M., {Dissauer}, K., {et~al.} 2015, \apj, 812,
  173, \dodoi{10.1088/0004-637X/812/2/173}

\bibitem[{{Wang} {et~al.}(2021){Wang}, {Chen}, \& {Ding}}]{wang21}
{Wang}, C., {Chen}, F., \& {Ding}, M. 2021, \apjl, 911, L8,
  \dodoi{10.3847/2041-8213/abefe6}

\bibitem[{{Wang} {et~al.}(2024){Wang}, {Ofman}, \& {Bradshaw}}]{wan24}
{Wang}, T., {Ofman}, L., \& {Bradshaw}, S.~J. 2024, \solphys, 299, 37,
  \dodoi{10.1007/s11207-024-02285-z}

\bibitem[{{Wang}(2016)}]{wan16}
{Wang}, T.~J. 2016, Washington DC American Geophysical Union Geophysical
  Monograph Series, 216, 395, \dodoi{10.1002/9781119055006.ch23}

\bibitem[{{Warmuth}(2015)}]{warm15}
{Warmuth}, A. 2015, Living Reviews in Solar Physics, 12, 3,
  \dodoi{10.1007/lrsp-2015-3}

\bibitem[{{Warren} {et~al.}(2018){Warren}, {Crump}, {Ugarte-Urra}, {Sun},
  {Aschwanden}, \& {Wiegelmann}}]{warr18}
{Warren}, H.~P., {Crump}, N.~A., {Ugarte-Urra}, I., {et~al.} 2018, \apj, 860,
  46, \dodoi{10.3847/1538-4357/aac20b}

\bibitem[{{Wiegelmann} \& {Sakurai}(2012)}]{wieg12}
{Wiegelmann}, T., \& {Sakurai}, T. 2012, Living Reviews in Solar Physics, 9, 5,
  \dodoi{10.12942/lrsp-2012-5}

\bibitem[{{Yang} {et~al.}(2015){Yang}, {Zhang}, {He}, {Peter}, {Tu}, {Wang},
  {Zhang}, \& {Feng}}]{yang15}
{Yang}, L., {Zhang}, L., {He}, J., {et~al.} 2015, \apj, 800, 111,
  \dodoi{10.1088/0004-637X/800/2/111}

\bibitem[{{Young} {et~al.}(2007){Young}, {Del Zanna}, {Mason}, {Dere}, {Landi},
  {Landini}, {Doschek}, {Brown}, {Culhane}, {Harra}, {Watanabe}, \&
  {Hara}}]{you07}
{Young}, P.~R., {Del Zanna}, G., {Mason}, H.~E., {et~al.} 2007, \pasj, 59,
  S857, \dodoi{10.1093/pasj/59.sp3.S857}

\bibitem[{{Zhang} {et~al.}(2022){Zhang}, {Chen}, {Liu}, \& {Wang}}]{zha22}
{Zhang}, P., {Chen}, J., {Liu}, R., \& {Wang}, C. 2022, \apj, 937, 26,
  \dodoi{10.3847/1538-4357/ac8d61}

\bibitem[{{Zhang} {et~al.}(2015){Zhang}, {Zhang}, {Wang}, \&
  {Nakariakov}}]{zhang15}
{Zhang}, Y., {Zhang}, J., {Wang}, J., \& {Nakariakov}, V.~M. 2015, \aap, 581,
  A78, \dodoi{10.1051/0004-6361/201525621}

\bibitem[{{Zheng}(2024)}]{zhen24}
{Zheng}, R. 2024, Proceedings of the Royal Society of London Series A, 480,
  20230950, \dodoi{10.1098/rspa.2023.0950}

\bibitem[{{Zhou} {et~al.}(2021){Zhou}, {Shen}, {Su}, {Tang}, {Zhou}, {Duan}, \&
  {Tan}}]{zhou21}
{Zhou}, X., {Shen}, Y., {Su}, J., {et~al.} 2021, \solphys, 296, 169,
  \dodoi{10.1007/s11207-021-01913-2}

\bibitem[{{Zimovets} {et~al.}(2021){Zimovets}, {McLaughlin}, {Srivastava},
  {Kolotkov}, {Kuznetsov}, {Kupriyanova}, {Cho}, {Inglis}, {Reale}, {Pascoe},
  {Tian}, {Yuan}, {Li}, \& {Zhang}}]{zim21}
{Zimovets}, I.~V., {McLaughlin}, J.~A., {Srivastava}, A.~K., {et~al.} 2021,
  \ssr, 217, 66, \dodoi{10.1007/s11214-021-00840-9}

\end{thebibliography}
\bibliographystyle{aasjournal}

\end{document}